\documentclass[10pt]{article}

\usepackage{amsbsy, amsfonts, amssymb, amsmath, dsfont}
\usepackage{paralist}
\usepackage{caption, graphicx, subfig, color}
\usepackage[text={6.75in,9.5in},centering,letterpaper]{geometry}
\usepackage[numbers, comma, square, sort&compress]{natbib}
\usepackage[hypertexnames=false, urlcolor=blue, colorlinks=true, linkcolor=blue, citecolor=blue]{hyperref}

\newtheorem{Lemma}{Lemma}[section]

\newtheorem{Hypothesis}{Hypothesis}

\newenvironment{Proof}[1][.]%
 {\begin{trivlist}\item[]\textbf{Proof#1 }}%
 {\hspace*{\fill}$\rule{0.3\baselineskip}{0.35\baselineskip}$\end{trivlist}}

\newenvironment{Acknowledgment}%
 {\begin{trivlist}\item[]\textbf{Acknowledgments.}}{\end{trivlist}}

\makeatletter\@addtoreset{equation}{section}\makeatother

\def\Prob{\mathop\mathrm{Prob}\nolimits}
\def\sign{\mathop\mathrm{sgn}\nolimits}
\def\ev{\mathop\mathrm{ev}\nolimits}
\def\argmax{\mathop\mathrm{arg~max}}

\newcommand{\N}{\mathbb{N}}             
\newcommand{\R}{\mathbb{R}}             
\newcommand{\Z}{\mathbb{Z}}             

\newcommand{\rmO}{\mathrm{O}}           
\newcommand{\rmd}{\mathrm{d}}           
\newcommand{\rme}{\mathrm{e}}           
\newcommand{\rmi}{\mathrm{i}}           

\newcommand{\changed}[1]{#1}

\begin{document}

\title{Data-driven continuation of patterns and their bifurcations}

\author{%
Wenjun Zhao\\
Division of Applied Mathematics\\
Brown University\\
Providence, RI~02912, USA
\and
Samuel Maffa\\
Broad Institute of MIT and Harvard\\
Cambridge, MA~02142, USA
\and
Bj\"orn Sandstede\\
Division of Applied Mathematics\\
Brown University\\
Providence, RI~02912, USA
}

\date{\today}
\maketitle

\begin{abstract}
Patterns and nonlinear waves, such as spots, stripes, and rotating spirals, arise prominently in many natural processes and in reaction-diffusion models. Our goal is to compute boundaries between parameter regions with different prevailing patterns and waves. We accomplish this by evolving randomized initial data to full patterns and evaluate feature functions, such as the number of connected components or their area distribution, on their sublevel sets. The resulting probability measure on the feature space, which we refer to as pattern statistics, can then be compared at different parameter values using the Wasserstein distance. We show that arclength predictor-corrector continuation can be used to trace out transition and bifurcation curves in parameter space by maximizing the distance of the pattern statistics. The utility of this approach is demonstrated through a range of examples involving homogeneous states, spots, stripes, and spiral waves.
\end{abstract}


\section{Introduction} \label{s:1}

\changed{Many biological, chemical, physical, and social systems exhibit spontaneous self-organization on microscopic scales that leads to coherent patterns and structures at macroscopic scales \cite{Busse, CrossHohenberg, Hoyle, Pismen}. Hexagonal cell structures, directional stripes, and spiral waves are examples of these patterns that arise, for instance, in cardiac tissue \cite{Cherry, Rosenbaum}, chemical reactions such as the chlorite-iodide-malonic acid reaction \cite{Boissonade, Ouyang}, and fluid flow such as the Rayleigh--B\'enard convection \cite{Bodenschatz}. Other examples are pigment patterns on zebrafish and other organisms \cite{zebrafish-tda}, vegetation patterns in semiarid environments \cite{Gowda}, schools and swarms formed by animals \cite{Bernoff, Katz}, and collective cell structures formed by homogeneous or heterogeneous cell populations \cite{Volkening, Buttenschon2020, Blanchard2019, Giniunaite2020, Osborne2017}.}

\changed{These systems often exhibit different spatial patterns that may occupy different regions in space, leading to the formation of interfaces between them that may, or may not, propagate. We say that a spatial pattern $A$ is prevalent if it invades other spatial patterns so that the region occupied by pattern $A$ increases in time until it fills the entire domain; in other words, its interface with other patterns moves into the region occupied by these other patterns. Depending on system parameters, different patterns may be prevalent in different parameter regions, and the boundaries of these parameter regions correspond to coexistence of two such patterns where the interface separating the spatial regions occupied by each pattern is stationary.}

\changed{We are interested in determining the specific region in parameter space where a given pattern is prevalent and delineating the bifurcation and transition curves that constitute the boundaries of these regions. This question is important in many applications. For instance, instabilities of spiral waves due to breakup or period doubling is relevant for the onset of cardiac arrhythmias in realistic models \cite{Cherry, Rosenbaum}, the transition between heat conduction and convection is relevant for mixing of fluid particles \cite{Bodenschatz}, different phenotypes of zebrafish can be linked to different gene mutations \cite{Volkening, zebrafish-tda}, and understanding different cell aggregation patterns may help with identifying how heterogeneous cell populations, for instance epithelial cell populations during wound healing, interact \cite{Bhaskar}.}

\changed{To explore the dynamics of a given system, identify which patterns it may exhibit, and construct a coarse-grained bifurcation diagram, a common strategy is to divide parameter space into a grid with a small spacing between adjacent nodes. At each grid node, direct numerical simulations starting from random initial conditions will likely produce the pattern that is prevalent at these parameter values. Comparing the resulting patterns by eye will then provide a coarse-grained bifurcation diagram. This method provides an excellent overview of emerging patterns and is relatively inexpensive, but it relies on visual inspection of the patterns emerging in the direct simulations, and it may produce inaccurate results in bistable regions where more than one pattern is stable.}

\changed{Accurate and efficient approaches for continuing patterns and their bifurcation as well as stationary or moving interfaces between two patterns exist in situations where the underlying model is a partial differential equation (PDE) and where these structures can be found as regular roots of an appropriate boundary-value problem. Solving this boundary-value problem using Newton's method and continuing these solutions using arclength continuation, as implemented, for instance, in Auto07p \cite{auto07p} or pde2path \cite{Uecker}, allows us to compute the corresponding bifurcation or transition curves in parameter space. This approach is highly accurate and very effective \citep{SurveyNumerics}, and it has been used successfully to compute spiral waves and interfaces between domain-filling patterns (see, for instance, \citep{Morrissey, Avery, Barkley, Dodson, Avitabile, Lloyd, LloydScheel, SandstedeScheel, SurveyCoherent} and references therein). The main disadvantage is that the boundary-value problem needs to be set up separately for each specific bifurcation or interface computation. While recent advantages in core-farfield decompositions \citep{Avery, Morrissey, LloydScheel} made this approach more applicable for interface computations, they remain difficult to implement. For instance, bifurcations can be traced out only when the underlying mechanism is known, and patterns can be continued only when we can identify the correct boundary-value problem formulation as a Fredholm system with index zero: there are many cases where this is difficult \cite{Avery, Morrissey, LloydScheel, Roberts}. In addition, domain-filling patterns typically exist for a range of wave numbers, and there are no known criteria to determine which wave number will be selected in a given system, even in the case of interfaces between wave trains with similar wave numbers \citep{Doelman, Rijk, Iyer, Johnson}. }

\begin{figure}
\centering
\includegraphics[scale=1.]{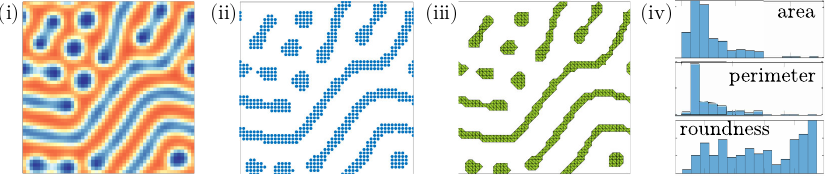}
\caption{The panels illustrate (i) a color plot of the solution of the underlying PDE model at time that is so large that spots and stripes have emerged, (ii) the corresponding sublevel set, (iii) the associated $\alpha$-shape (a polygonal approximation of the boundary of the sublevel set), and (iv) the distribution (histogram) of three different feature functions evaluated on each of ten simulations that start from different random initial data, namely the areas of the connected components of the $\alpha$-shape (top), their perimeters (center), and the roundness score, which is the inverse of the isoperimetric ratio or, equivalently, the fraction of area over perimeter and therefore measures of how elongated each connected component is (bottom).}
\label{fig:illustration_feature}
\end{figure}

\changed{In this paper, we develop an alternative framework based on pattern statistics to characterize and distinguish different patterns such as spots and stripes, different bifurcation and transition curves, and different dynamical behaviors such as rotating, meandering, or turbulent spiral waves. We justify this approach by providing a mathematically rigorous foundation for reaction-diffusion system. We apply the resulting continuation algorithm primarily to paradigm reaction-diffusion systems as this allows us to compare our results with known numerical and theoretical results. We note that our approach is not limited to PDEs but can also be applied to stochastic agent-based models such as those considered or reviewed in \cite{Bullara, Volkening, Buttenschon2020, Blanchard2019, Giniunaite2020, Osborne2017, zebrafish-tda, Bhaskar}, where the observed patterns are often noisy and change from simulation to simulation. We demonstrate the utility of our algorithm for tracing out transition curves in stochastic systems by applying it to a lattice model of zebrafish patterning.}

\begin{figure}
\centering
\includegraphics[width=0.9\textwidth]{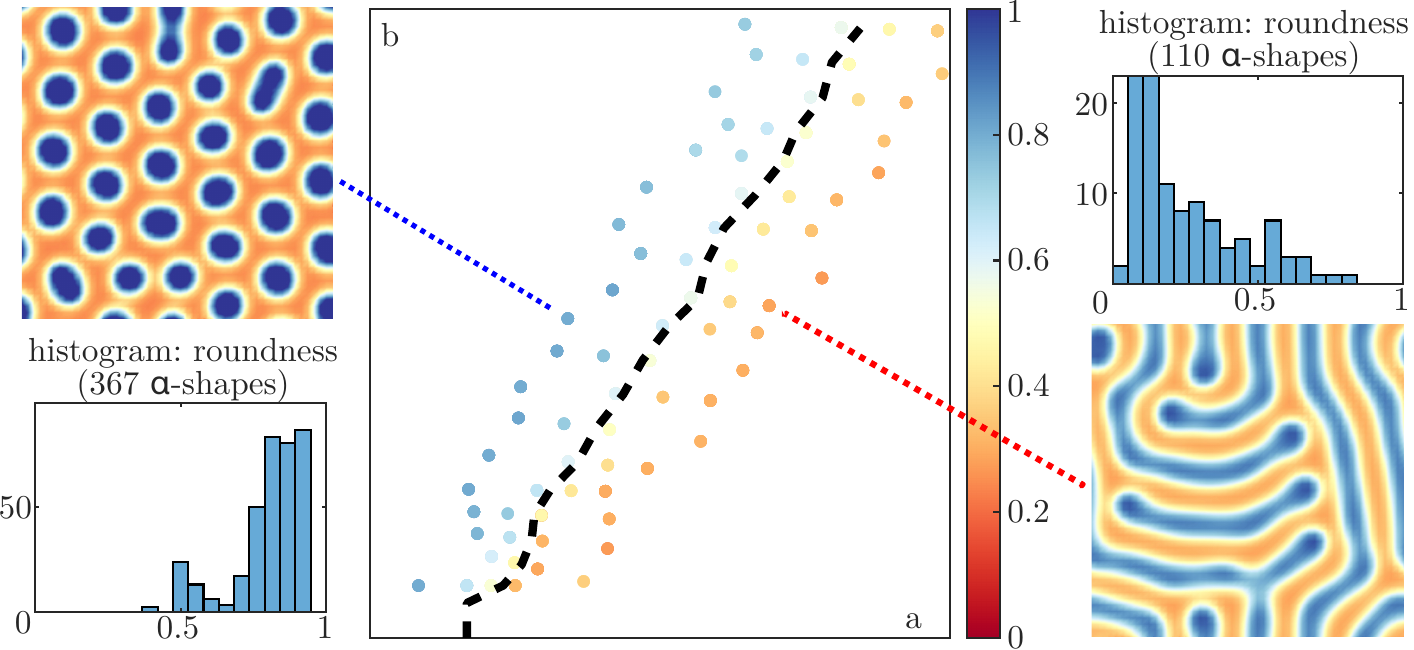}
\caption{Shown is a segment of the transition curve that separates the regions in parameter space where, respectively, spots and stripes are prevalent in the Brusselator model. The feature function is given by the roundness scores of the connected components. Sample pattern statistics (given by histograms of feature evaluations of direct simulations with \changed{an ensemble of} randomized initial data) and sample patterns are included in the insets. The colored disks correspond to parameter values where the pattern statistics was computed during continuation, with colors indicating the expectation of the roundness score.}
\label{fig:illustration}
\end{figure}

\changed{Our proposed scheme relies on direct numerical simulations as these are inexpensive, relatively easy to implement, and available for PDEs and stochastic agent-based models alike.} To use direct simulations within a continuation framework, we need to be able to characterize and differentiate the patterns we try to distinguish. One option is to use topological data analysis as was done in \citep{Shipman, harrington} for domain-filling hexagon patterns or in \citep{zebrafish-tda, zebrafish-comparison, Bhaskar} for heterogeneous cell populations arising in agent-based models. More generally, we can evolve \changed{an ensemble of initial conditions} until a fixed time that is chosen so large that patterns have emerged. We then consider an appropriate sublevel set of the solution and evaluate a feature function on the sublevel set. Examples of feature functions are the number of connected components of the sublevel set, their area distribution, or the distribution of their roundness scores (which is the ratio of area and perimeter of a connected component \changed{or, equivalently, the inverse of the isoperimetric ratio}, thus reflecting its elongation); see Figure~\ref{fig:illustration_feature}  for an illustration. In practice, sublevel sets are approximated by $\alpha$-shapes, which are polygonal approximations of the boundary of the sublevel sets \citep{Edelsbrunner}. We consider feature functions that map into a metric space $\mathcal{Z}$ and conduct several simulations starting from randomized initial data to obtain an empirical probability measure on $\mathcal{Z}$, which we refer to as the \emph{pattern statistics}. We quantify the difference between two pattern statistics via the Wasserstein distance. We choose feature functions tailored to specific bifurcations or transitions: for instance, the number of connected components distinguishes spots and stripes, and the area of the region traced out by the tip of a spiral waves differentiates rotating from meandering spirals. Maximizing the Wasserstein distance between the pattern statistics computed at two nearby points in parameter space allows us to compute the corresponding transition curve using predictor-corrector arclength continuation; see Figure~\ref{fig:illustration} for a result for the Brusselator model. The proposed methodology is purely data-driven and enables automated and efficient bifurcation tracing with limited prior knowledge of the underlying system.

Our manuscript is organized as follows. In the next section, we will introduce a probabilistic framework that serves as a theoretical foundation for pattern statistics and their computation. In particular, we will show that the pipeline from choosing an ensemble of initial data, evolving these for a fixed time $T$, mapping to the sublevel set $u^{-1}((-\infty,c])$ for a fixed value of $c$, and evaluating a feature function on the resulting set with values in a compact metric space $\mathcal{Z}$ will, under appropriate assumptions, yield a probability measure on the feature space $\mathcal{Z}$, our \emph{pattern statistics}, that depends continuously on parameters in the Wasserstein metric. Afterwards, we discuss how we can use pattern statistics to compute bifurcation and transition curves based on predictor-corrector algorithms that utilize bisection and quadratic interpolation, \changed{and we also discuss the accuracy, computational cost, and robustness under changes of the computational parameters of our algorithm}. Finally, we will demonstrate the utility and applicability of the proposed continuation framework through a range of examples involving homogeneous states, spots, stripes, and spiral waves. Specifically, we will show that the approach can be used to trace out
\begin{compactitem}
\item curves corresponding to stationary interfaces between homogeneous states, spots, and stripe patterns,
\item fold bifurcations of spots and stripes,
\item boundaries of snaking regions, and
\item transitions between rigidly-rotating, meandering, drifting, period-doubled, and turbulent spiral waves.
\end{compactitem}
We will also outline how our approach via $\alpha$-shapes can be used to compute spiral waves via the \emph{freezing method} \citep{Thuemmler1,Thuemmler2} without the need to implement solvers for algebraic-differential systems. Finally, we will describe other potential applications and extensions to multi-parameter continuation in the discussion section.

\begin{Acknowledgment}
Sandstede was partially supported by the NSF under grants DMS-2038039 and DMS-2106566.
\end{Acknowledgment}


\section{Pattern statistics: A probabilistic framework} \label{s:2}

\subsubsection*{Reaction-diffusion models}

We consider nonlinear reaction-diffusion systems on bounded square domains with periodic boundary conditions of the form
\begin{equation}\label{e:pde}
\frac{\partial U}{\partial t} = \mathcal{D} \Delta U + \mathcal{N}(U,p), \quad
x\in D:=(\R/2\pi\Z)^2, \quad U\in\R^d, \quad p\in\mathcal{P}\subset\R^2.
\end{equation}
We note that our approach applies also to other bounded domains, but we will not consider these for simplicity. We are interested in domains that are large compared to the typical wavenumber $\kappa$ of patterns exhibited by (\ref{e:pde}) and therefore assume that $0<\kappa\ll1$. We also assume that the parameter region $\mathcal{P}$ is compact with open interior. We denote the nonlinear semiflow associated with (\ref{e:pde}) by $\Phi_t(U_0,p)$: standard regularity theory allows us to view $\Phi_t$ for each $t>0$ as a smooth map from $H^2(D,\R^d)\times\mathcal{P}$ into $C^2(D,\R^d)$.

\subsubsection*{Spaces of probability measures}

We begin by briefly reviewing push-forward measures and spaces of probability measures equipped with the 2-Wasserstein distance. If $(\Omega,\sigma)$ and $(\tilde\Omega,\tilde\sigma)$ are measurable spaces, $F\colon\Omega\to\tilde\Omega$ is measurable, and $\mu$ is a probability measure on $(\Omega,\sigma)$, then the push-forward $F_\#\mu$, defined by $F_\#\mu(B):=\mu(F^{-1}(B))$ for all $B\in\tilde\sigma$, is a probability measure on $(\tilde\Omega,\tilde\sigma)$. From now on, let $(\mathcal{Z},d_\mathcal{Z})$ be a compact metric space equipped with the Borel $\sigma$-algebra. If $F\colon\Omega\to\mathcal{Z}$ is measurable and $g\in C^0(\mathcal{Z},\R)$, then we have
\begin{equation}\label{d:it}
\int_\mathcal{Z} g\,\rmd F_\#\mu = \int_\Omega g\circ F\,\rmd\mu.
\end{equation}
We denote by $\Prob(\mathcal{Z})$ the space of probability measures on $\mathcal{Z}$ equipped with the Borel $\sigma$-algebra. We say that a sequence $(\mu_n)_{n\in\N}$ of measures in $\Prob(\mathcal{Z})$ converges weakly to $\mu\in\Prob(\mathcal{Z})$, denoted by $\mu_n\rightharpoonup\mu$, if for each $g\in C^0(\mathcal{Z},\R)$ we have
\begin{equation}\label{d:wc}
\int_\mathcal{Z} g\,\rmd\mu_n \longrightarrow \int_\mathcal{Z} g\,\rmd\mu \quad\mbox{ as } n\to\infty.
\end{equation}
Given $\mu,\nu\in\Prob(\mathcal{Z})$, we denote by $\Pi(\mu,\nu):=\{\pi\in\Prob(\mathcal{Z}\times\mathcal{Z})\colon (P_1)_\#\pi=\mu,\; (P_2)_\#\pi=\nu\}$ the space of couplings of $(\mu,\nu)$, that is, the space of probability measures on $\mathcal{Z}\times\mathcal{Z}$ with marginals $\mu$ and $\nu$, where $P_j\colon \mathcal{Z}\times\mathcal{Z}\to\mathcal{Z}$ projects onto the $j$th component. We equip $\Prob(\mathcal{Z})$ with the 2-Wasserstein distance defined by
\begin{equation}\label{e:W2}
d_\mathrm{W}(\mu,\nu) := \left( \inf_{\pi\in\Pi(\mu,\nu)} \int_{\mathcal{Z}\times \mathcal{Z}} d_\mathcal{Z}(x,y)^2\, \rmd\pi(x,y) \right)^{\frac12}, \qquad
\mu,\nu\in\Prob(\mathcal{Z}).
\end{equation}
The function $d_\mathrm{W}$ is a metric on $\Prob(\mathcal{Z})$ that metrizes the weak convergence of measures on $\mathcal{Z}$ \citep[Corollary~6.13]{Villani} (that is, $\mu_n\rightharpoonup\mu$ if and only if $d_\mathrm{W}(\mu_n,\mu)\to0$), and the metric space $(\Prob(\mathcal{Z}),d_\mathrm{W})$ is compact \citep[Theorem~5.1.3, Equation (5.1.20), and Proposition~7.1.5]{Ambrosio}. For empirical measures of the form $\mu=\frac{1}{N}\sum_{n=1}^N\delta_{z_n}$ and $\nu=\frac{1}{N}\sum_{n=1}^N\delta_{\tilde{z}_n}$ with $z_n,\tilde{z}_n\in\mathcal{\mathcal{Z}}$, the Wasserstein distance becomes the discrete optimal-transport problem
\begin{equation}\label{e:wot}
d_\mathrm{W}(\mu,\nu) = \left( \inf\left\{\sum_{m,n=1}^N \Gamma_{mn}d_\mathcal{Z}(z_m,\tilde{z}_n)^2\colon\Gamma\in\R^{N\times N},\, \Gamma_{mn}\geq0,\, \sum_{j=1}^N \Gamma_{jn}=\sum_{j=1}^N \Gamma_{mj}=\frac{1}{N}\, \forall m,n\right\}\right)^{\frac12},
\end{equation}
which for the case $\mathcal{Z}=I\subset\R$ gives
\begin{equation}\label{e:WEmp}
d_\mathrm{W}(\mu,\nu) = \changed{\left(\frac{1}{N} \sum_{n=1}^N |z_n-\tilde{z}_n|^2\right)^{\frac12}}
\mbox{ assuming the ordering } z_1\leq\ldots\leq z_N,\; \tilde{z}_1\leq\ldots\leq\tilde{z}_N;
\end{equation}
see \cite[Remark~2.28]{Peyre}. Finally, when $\mathcal{Z}=I\subset\R$, the expectation
\begin{equation}\label{e:Exp}
E\colon\Prob(I)\longrightarrow I,\quad \mu\longmapsto E(\mu):=\int_I z\,\rmd\mu(z)
\end{equation}
is continuous in the Wasserstein distance.

\subsubsection*{Sublevel sets and patterns}

We focus on feature functions that operate on closed sublevel sets of, say, the first component $U_1$ of $U\in C^2(D,\R^d)$, that is on sets of the form $\{x\in D\colon U_1(x)\leq c\}$ for some threshold $c$; we note that the same results apply also to sets of the form $\{x\in D\colon U_1(x)\geq c\}$. To formalize our notation, choose a threshold $c\in\R$, let $\mathcal{X}:=C^2(D,\R)$, and define
\begin{equation}\label{e:Xr}
\mathcal{X}_\mathrm{reg} := \left\{ u\in C^2(D,\R)\colon u^{-1}(c)\neq\emptyset, \mbox{ and } \nabla u(x)\neq0 \mbox{ for all } x\in u^{-1}(c) \right\}
\end{equation}
to be the set of functions in $C^2(D,\R)$ that have $c$ as a regular value and attain this value. Since $D$ is compact, we know that $\mathcal{X}_\mathrm{reg}$ is open in $\mathcal{X}$. Throughout, we refer to sublevel sets of the form $u^{-1}((-\infty,c])$ for a function $u\in\mathcal{X}_\mathrm{reg}$ as a \emph{pattern}.

For completeness, we first state a result on the existence of $C^2$ tubular neighborhoods of compact one-dimensional compact $C^2$ submanifolds (or 1-manifolds, for short) of the torus. Note that each 1-manifold inside the torus is the finite disjoint union of one-dimensional manifolds that are each $C^2$-diffeomorphic to a circle.

\begin{Lemma}\label{ln:1}
Let $B$ be a one-dimensional compact $C^2$ submanifold of $D$, then there exists an open neighborhood $V$ of $B$ in $D$ and a $C^2$-diffeomorphism $\theta\colon B\times(-1,1)\to V$, $(b,y)\mapsto\theta(b,y)$ so that $\theta|_{B\times\{0\}}$ is a $C^2$-diffeomorphism onto $B$. The map $\theta$ is referred to as a tubular neighborhood of $B$ in $D$.
\end{Lemma}

\begin{Proof}
From \cite[Theorem~3.6 in Chapter~2]{Hirsch}, we know that there is a $C^\infty$ manifold pair $(D^\infty,B^\infty)$ with $D^\infty\subset\R^3$ and a $C^2$-diffeomorphism $\theta_1\colon(D^\infty,B^\infty)\to(D,B)$. Next, \cite[Theorem~5.2 in Chapter~4]{Hirsch} shows that $B^\infty$ has a $C^\infty$ tubular neighborhood $\theta_2$ in $D^\infty$ so that $\theta_2\colon B^\infty\times(-1,1) \to D^\infty$ is a $C^\infty$ diffeomorphism onto an open neighborhood of $B^\infty$ in $D^\infty$ and $\theta_2|_{B^\infty\times\{0\}}$ is a diffeomorphism onto $B^\infty$. Hence, $\theta:=\theta_2\circ\theta_1$ is a $C^2$ tubular neighborhood of $B$ in $D$.
\end{Proof}

For $u\in\mathcal{X}_\mathrm{reg}$, we denote by $A(u):=u^{-1}((-\infty,c])$ the sublevel set of $u$. Our next result shows that $A(u)$ is a $C^2$-submanifold with boundary $\partial A(u)=u^{-1}(c)$ and describes how $A(u)$ changes as $u$ varies in $\mathcal{X}_\mathrm{reg}$.

\begin{Lemma}\label{ln:2}
For each $u_0\in\mathcal{X}_\mathrm{reg}$, the sublevel set $A_0:=A(u_0)=u_0^{-1}((-\infty,c])$ is a $C^2$ submanifold of $D$ with boundary $\partial A_0=u_0^{-1}(c)$. Furthermore, there are open neighborhoods $V\subset D$ and $\mathcal{U}\subset C^2(D,\R)$ of $\partial A_0$ and $u_0$, respectively, and a $C^2$ map $\tau\colon A_0\times\mathcal{U}\to D$ so that for each $u\in\mathcal{U}$ the map $\tau(\cdot,u)\colon A_0\to D$ is a $C^2$-diffeomorphism from $(A_0,\partial A_0)$ onto $(A(u),\partial A(u))$ with $\tau(a,u)=a$ for all $a\in A_0\setminus V$ and $\tau(a,u_0)=a$ for all $a\in A_0$.
\end{Lemma}

\begin{Proof}
Since $u_0\in\mathcal{X}_\mathrm{reg}$, we have $u_0^{-1}(c)\neq\emptyset$ and $\nabla u_0(b)\neq0$ for each $b\in u_0^{-1}(c)$. Hence, $u_0^{-1}(c)$ is a nonempty $C^2$ 1-manifold in $D$. Furthermore, $u_0(a)$ assumes all values near $c$ for appropriate values of $a\in D$ near $b\in u_0^{-1}(c)$, and we conclude that $u_0^{-1}((-\infty,c])$ is a $C^2$ submanifold with boundary $\partial A_0=u_0^{-1}(c)$ as claimed.

Next, let $I=(-1,1)$ and denote by $\theta\colon\partial A_0\times I\to V$ a $C^2$ tubular neighborhood of $\partial A_0$ in $D$. Since $\nabla u_0(b)\neq0$ for all $b\in u_0^{-1}(c)$, we can choose $\theta$ so that $\theta(\partial A_0\times(-1,0])=A_0\cap V$.

The evaluation map $\ev\colon\partial A_0\times I\times\R\times C^2(D,\R)\to\R$ given by $\ev(b,y,\alpha,u):=u(h(b,y))-\alpha\in\R$ is $C^2$ by \cite[Theorem~10.10 in Chapter~2]{ChowHale} or \cite[Corollary~11.7]{Michor}. By definition, we have $\ev(b,0,c,u_0)=0$ with $\ev_\alpha(b,0,c,u_0)=-1$ for all $b\in\partial A_0$, and there is a $\delta_0>0$ so that $|\ev_y(b,0,c,u_0)|\geq\delta_0$ for $b\in \partial A_0$ since $\partial A_0$ is compact and $\nabla u_0(b)\neq0$ for all $b\in\partial A_0$. Hence, we can apply the implicit function theorem to the equation $\ev(b,y,\alpha,u)=0$ near each point $(b,0,c,u_0)$ with $b\in\partial A_0$ to obtain the existence of a neighborhood $\mathcal{U}$ of $u_0$ in $C^2(D,\R)$, open intervals $I,J\subset\R$ with $0\in I$ and $c\in J$, and unique $C^2$ maps $\psi_0\colon\partial A_0\times J\times\mathcal{U}\to I$ and $\psi_1\colon\partial A_0\times I\times\mathcal{U}\to J$ so that $u(h(b,y))=\alpha$ for $(b,y,\alpha,u)\in\partial A_0\times I\times J\times\mathcal{U}$ if and only if $y=\psi_0(b,\alpha,u)$ or, equivalently, $\alpha=\psi_1(b,y,u)$. In particular, the maps $\psi_0(b,\cdot,u)$ and $\psi_1(b,\cdot,u)$ are inverses of each other for each fixed $(b,u)$, and we have $\psi_0(\cdot,c,u_0)\equiv0$ and $\psi_1(\cdot,0,u_0)\equiv c$.

Let $\chi\colon J\to[0,1]$ be a $C^\infty$ cutoff function with $\chi(\alpha)=0$ for $\alpha$ near $c$ and $\chi(\alpha)=1$ for $\alpha$ near $\partial J$. We define
\[
\tilde\tau\colon\partial A_0\times I\times\mathcal{U}\longrightarrow\partial A_0\times I, \quad
(b,y,u) \longmapsto (b,\psi_1(b,y,u_0),u) =: (b,\alpha,u) \longmapsto
\left( b,\psi_0(b,\alpha,u+\chi(\alpha)(u_0-u)) \right) 
\]
so that $\tilde\tau$ is a $C^2$ diffeomorphism with $\tilde\tau(\partial A_0\times\{0\},u)=\theta^{-1}(\partial A(u))$ and $\tilde\tau(\partial A_0\times(-1,0],u)=\theta^{-1}(A(u)\cap V)$. For each $u$, we define the diffeomorphism $\tau(\cdot,u)$ by $\tau(\cdot,u):=\theta\circ\tilde\tau(\cdot,u)\circ\theta^{-1}$ on $A_0\cap V$ and extend it to $A_0\setminus V$ by the identity.
\end{Proof}

Next, we will use the characterization of $A(u)=u^{-1}((-\infty,c])$ provided in Lemma~\ref{ln:2} to analyze feature functions.

\subsubsection*{Feature functions}

The intuition from Lemma~\ref{ln:2} is that each function that operates continuously on finite disjoint unions of connected two-dimensional $C^2$ manifolds with boundaries inside the torus $D$ defines a feature function. Our goal is to formalize this notion, give a few examples of feature functions that will be used later to distinguish spatially homogeneous states, stripe patterns, and spot patterns, and prove that these feature functions satisfy our formal definition.

We first formalize our notion of feature functions. Let $\mathcal{Z}$ be a compact metric space. We say that a function $f\colon\mathcal{X}_\mathrm{reg}\to\mathcal{Z}$ is a \emph{feature function} provided $f$ is continuous. Since $\mathcal{X}_\mathrm{reg}$ is open in $C^2(D,\R)$, we can extend each feature function $f\colon\mathcal{X}_\mathrm{reg}\to\mathcal{Z}$ to a measurable function $f\colon C^2(D,\R)\to\mathcal{Z}$ by mapping $C^2(D,\R)\setminus\mathcal{X}_\mathrm{reg}$ to an arbitrary fixed element in $\mathcal{Z}$.

We now introduce several functions that map $\mathcal{X}_\mathrm{reg}$ into appropriate compact metric spaces $\mathcal{Z}$. For each two-dimensional $C^2$ submanifold $A$ with boundary of the torus $D$, we denote by $\mu_\mathrm{Leb}(A)$ its Lebesgue measure. To ensure compactness of the range of some of the feature functions we introduce, we choose an $\mathfrak{m}\gg1$ and cap some of the quantities below at $\mathfrak{m}$. We will again use the notation $A(u):=u^{-1}((-\infty,c])$ for elements $u\in\mathcal{X}_\mathrm{reg}$. With this notation, we define the following functions:
\begin{compactenum}[(1)]
\item\label{lconn}
$f_\mathrm{Conn}(u):=\min\{\beta(A(u)),\mathfrak{m}\}\in\mathcal{Z}_\mathrm{Conn}:=\N\cap[0,\mathfrak{m}]$ is the zeroth Betti number $\beta(A(u))$, that is, the number of connected components of $A(u)$ (capped at $\mathfrak{m}$).
\item\label{lleb}
$f_\mathrm{Leb}(u):=\mu_\mathrm{Leb}(A(u))\in\mathcal{Z}_\mathrm{Leb}:=[0,\mu_\mathrm{Leb}(D)]$ is the Lebesgue measure of $A(u)$.
\item\label{larea}
$f_\mathrm{AreaDistr}(u):=\frac{1}{\beta(A(u))}\sum_{j=1}^{\beta(A(u))}\delta_{\mu_\mathrm{Leb}(A_j(u))}\changed{\in\mathcal{Z}_\mathrm{AreaDistr}}$ is the probability measure with atoms on the areas of the connected components $A_j(u)$ of $A(u)$, where $\mathcal{Z}_\mathrm{AreaDistr}:=\Prob([0,\mu_\mathrm{Leb}(D)])$ is equipped with the 2-Wasserstein distance.
\item\label{lroundness}
$f_\mathrm{RoundDistr}(u):=\frac{1}{\beta(A(u))}\sum_{j=1}^{\beta(A(u))}\delta_{g(A_j(u))}\changed{\in\mathcal{Z}_\mathrm{RoundDistr}}$ is the probability measure with atoms on the roundness scores
\[
g(A_j(u)) := \min\left\{ \frac{4\pi\mu_\mathrm{Leb}(A_j(u))}{|\mbox{Perimeter of }A_j(u)|^2}, \mathfrak{m} \right\} \in [0,\mathfrak{m}]
\]
of the connected components $A_j(u)$ of $A(u)$ (capped at $\mathfrak{m}$), where $\mathcal{Z}_\mathrm{RoundDistr}:= \Prob([0,\mathfrak{m}])$ is equipped with the 2-Wasserstein distance.
\end{compactenum}

For each connected $C^2$ submanifold $A$ with boundary in $D$, the individual roundness score $g(A)$ will be close to zero when $A$ is an elongated stripe, while it will be close to one when $A$ is close to a regular disk. The next lemma shows that the functions defined in (\ref{lconn})-(\ref{lroundness}) are indeed feature functions.

\begin{Lemma}\label{ln:3}
The functions defined in (\ref{lconn})-(\ref{lroundness}) are continuous from $\mathcal{X}_\mathrm{reg}$ into their respective ranges and therefore define feature functions.
\end{Lemma}

\begin{Proof}
We established in Lemma~\ref{ln:2} that $(A(u),\partial A(u))$ is diffeomorphic to $(A(u_0),\partial A(u_0))$ for all $u$ close to $u_0$ in $\mathcal{X}_\mathrm{reg}$ via the diffeomorphism $\tau(\cdot,u)$ with $\tau(\cdot,u_0)=\mathrm{id}$. In particular, the number $f_\mathrm{Conn}(u)$ of connected components of $A(u)$ is locally constant in $\mathcal{X}_\mathrm{reg}$, and therefore continuous. Similarly, we can combine the diffeomorphism $\tau(\cdot,u)$ with the transformation formulas for path and area integrals to conclude that the perimeter and area of each connected component of $A(u)$ depend continuously on $u$. Finally, equation (\ref{e:WEmp}) shows that the map that associated a vector $z=(z_1,\ldots,z_N)\in\R^N$ to the discrete probability measure $\frac{1}{N}\sum_{j=1}^N\delta_{z_j}$ is continuous in the 2-Wasserstein metric. This completes the proof.
\end{Proof}

If $\mathcal{Z}=\Prob(I)$ for a bounded interval $I\subset\R$, and $f\colon\mathcal{X}_\mathrm{reg}\to\mathcal{Z}$ is a feature function, then its composition $E\circ f\colon\mathcal{X}_\mathrm{reg}\to I$ with the expectation defined in (\ref{e:Exp}) is also a feature function. In particular, we can define the feature functions $f_\mathrm{E_{AreaDistr}}:=E\circ f_\mathrm{AreaDistr}$ and $f_\mathrm{E_{RoundDistr}}:=E\circ f_\mathrm{RoundDistr}$ which map a pattern to the expectation of, respectively, the area and roundness score distributions of its connected components.

\subsubsection*{Randomization of initial data}

Our next step is to construct \changed{ensembles of initial data that generate ensembles of solutions on which we can evaluate a given feature function. We accomplish this through randomization of initial data as in \citep{Burq,Nahmod}.} Recall that we pose the reaction-diffusion system (\ref{e:pde}) on the torus $D=(\R/2\pi\Z)^2$. From now on, we use the notation $H^2(D):=H^2(D,\R^d)$ and $\ell^2(\Z^2):=\ell^2(\Z^2,\R^d)$. We represent initial conditions $U\in H^2(D)$ via their Fourier series. Let $e_k(x)=\frac{1}{1+|k|^2}\rme^{\rmi\langle k,x\rangle}$ with $k\in\Z^2$ be the standard orthonormal Fourier-series basis of $H^2(D)$.

First, we choose two functions $U_\mathrm{r},U_\mathrm{b}\in H^2(D)$: the base function $U_\mathrm{b}$ will represent the mean of the randomized initial conditions, while $U_\mathrm{r}$ will be used for the actual randomization. We write $U_\mathrm{r}=\sum_{k\in\Z^2} a_k e_k(x)\in H^2(D)$ so that $(a_k)_{k\in\Z^2}\in\ell^2(\Z^2)$ satisfies $|U_\mathrm{r}|_{H^2(D)}=|(a_k)_k|_{\ell^2(\Z^2)}$.

Following \citep{Burq,Nahmod}, we now randomize the fixed function $U_\mathrm{r}$. We choose a probability space $(\Omega,\sigma,\mu_\Omega)$ and a sequence of independent, zero-mean $\R^d$-valued random variables $(b_k)_{k\in\Z^2}$ with the property that there is a constant $C_0>0$ so that
\begin{equation}\label{e:b}
\int_\Omega |b_k(\omega)|^2\,\rmd\mu_\Omega \leq C_0 \mbox{ for all } k\in\Z^2.
\end{equation}
We define
\begin{equation}\label{e:xi}
\xi\colon\quad \Omega \longrightarrow \ell^2(\Z^2), \quad
\omega \longmapsto (a_k \odot b_k(\omega))_{k\in\Z^2},
\end{equation}
where $\odot$ denotes elementwise multiplication of vectors in $\R^d$ (that is, for $a,b\in\R^d$ we set $a\odot b:=(a_jb_j)_{j=1,\ldots,d}\in\R^d$). Our first result shows that $\xi$ is a well-defined measurable function and therefore defines an $\ell^2$-valued random variable.

\begin{Lemma}\label{l:3}
The map $\xi$ defined in (\ref{e:xi}) is measurable with $\xi\in L^2(\Omega,\ell^2(\Z^2))$.
\end{Lemma}

\begin{Proof}
For each $N\geq1$, we define $\xi_N\colon\Omega\to\ell^2$ by $[\xi_N(\omega)]_k=\mathds{1}_{|k|\leq N}a_k\odot b_k(\omega)$ so that $[\xi_N(\omega)]_k=0$ for all $|k|>N$. These functions are measurable with $\xi_N\in L^2(\Omega,\ell^2)$ for each $N$. Using (\ref{e:b}), we have for each $M\geq N$ that
\[
\int_\Omega |\xi_M(\omega)-\xi_N(\omega)|_{\ell^2}^2 \,\rmd\mu_\Omega \leq
\sum_{k=N+1}^M \int_\Omega |a_k\odot b_k(\omega)|^2 \,\rmd\mu_\Omega \leq
\sum_{k=N+1}^M |a_k|^2 \int_\Omega |b_k(\omega)|^2 \,\rmd\mu_\Omega \leq
C_0 \sum_{k=N+1}^M |a_k|^2 \to 0
\]
as $M,N\to\infty$. Hence, $\xi_N\in L^2(\Omega,\ell^2)$ is a Cauchy sequence, and we conclude that there is a function $\xi\in L^2(\Omega,\ell^2)$ so that $\xi_N\to\xi$ in $L^2$ as $N\to\infty$. Furthermore, there is a subsequence $(N_n)_n$ so that $\xi_{N_n}(\omega)\to\xi(\omega)$ as $n\to\infty$ for almost every $\omega\in\Omega$, which shows that $\xi$ is given by (\ref{e:xi}) almost everywhere.
\end{Proof}

We now use the linear isomorphism $\iota\colon\ell^2(\Z^2)\to H^2(D)$ provided by the Fourier-series expansion to define the map
\begin{equation}\label{e:iota}
\iota\circ\xi\colon\quad \Omega \longrightarrow H^2(D), \quad
\omega \longmapsto U^\omega := U_\mathrm{b} + U_\mathrm{r}^\omega \quad\mbox{ with }
U_\mathrm{r}^\omega := \sum_{k\in\Z^2} (a_k \odot b_k(\omega)) e_k(x).
\end{equation}
Note that $\iota\circ\xi$ is measurable and defines a $H^2(D)$-valued random variable that lies in $L^2(\Omega,H^2(D))$.

Finally, we remark that some of the transition curves we discuss below can be computed efficiently using a single deterministic initial condition $U_\mathrm{b}$ instead of a randomization of initial data. These cases still fall into the framework discussed above upon defining $\Omega$ to consist of a single point and $\xi$ to map this point to $U_\mathrm{b}$.

\subsubsection*{Ensembles of solutions} 

Next, we propagate the randomized initial conditions forward in time to create ensembles of solutions that are parametrized by $\omega\in\Omega$. We fix a time $T>0$ and note that the semiflow $\Phi_T(U_0,p)$ of the reaction-diffusion system (\ref{e:pde}) is a smooth map from $H^2(D)\times\mathcal{P}$ into $C^2(D)$. We denote by $P_1\colon C^2(D)\to\mathcal{X}=C^2(D,\R)$, $U=(U_1,\ldots,U_d)\mapsto U_1$ the continuous projection onto the first component. The \emph{ensemble function}
\[
\mathcal{E}_p:=P_1\circ\Phi_T(\cdot,p)\circ\iota\colon\ell^2(\Z^2)\to\mathcal{X}
\]
then maps initial data in Fourier space to the first component of the solution evaluated at time $T$, and we see that the composite map
\begin{equation}
\mathcal{E}_p\circ\xi \colon \quad
\Omega \longrightarrow \mathcal{X} = C^2(D,\R), \quad
\omega \longmapsto P_1\Phi_T(U^\omega,p)
\end{equation}
is measurable for each $p\in\mathcal{P}$.

Recall that feature functions are assumed to be continuous only on the subset $\mathcal{X}_\mathrm{reg}$ of functions in $C^2(D,\R)$ that attain $c$ as a regular value. We will therefore assume that our randomization (consisting of our choices of $U_\mathrm{b},U_\mathrm{r}\in H^2(D)$ and the random variables $(b_k)_{k\in\Z^2}$), the constant $c$ appearing in the definition (\ref{e:Xr}) of $\mathcal{X}_\mathrm{reg}$, and the time $T>0$ can be chosen so that $\mathcal{E}_p\circ\xi$ maps almost surely into $\mathcal{X}_\mathrm{reg}$ for each $p\in\mathcal{P}$. We formalize this equivalently as follows. Since $\mathcal{E}_p\circ\xi$ is measurable for each $p$, the push-forward $\mu_\mathcal{X}(p):=(\mathcal{E}_p\circ\xi)_\#\mu_\Omega$ is a well-defined probability measure on $\mathcal{X}=C^2(D,\R)$ equipped with the Borel $\sigma$-algebra. We assume that the following hypothesis is met:

\begin{Hypothesis}\label{h1}
Assume that $(b_k)_{k\in\Z^2}$ are independent, zero-mean $\R^d$-valued random variables that satisfy (\ref{e:b}). Furthermore, assume that these random variables, the functions $U_\mathrm{b},U_\mathrm{r}\in H^2(D)$, and the constants $c\in\R$ and $T>0$ are such that $\mu_\mathcal{X}(p)(\mathcal{X}_\mathrm{reg})=1$ for each $p\in\mathcal{P}$.
\end{Hypothesis}

Our hypothesis essentially assumes that the set of $\omega$ for which the sublevel set of the solution at time $T$ is not a manifold (and instead undergoes a bifurcation) has measure zero. For fixed $p$, these bifurcations should occur at most along codimension-one sets, so the hypothesis should be satisfied for generic systems (and generic choices of the quantities mentioned in \ref{h1}). We will assume from now on that \ref{h1} is met.

\subsubsection*{Pattern statistics}

For each given feature function $f\colon\mathcal{X}_\mathrm{reg}\to\mathcal{Z}$, where $\mathcal{Z}$ is a compact metric space, the composition
\[
f\circ\mathcal{E}_p\circ\xi\colon\Omega\longrightarrow\mathcal{Z}
\]
is measurable. We define the \emph{pattern statistics} to be the map
\begin{equation}\label{e:featdistr}
\mu_f\colon\quad \mathcal{P} \longrightarrow \Prob(\mathcal{Z}), \quad
p \longmapsto \mu_f(p) := (f\circ\mathcal{E}_p\circ\xi)_\#\mu_\Omega
\end{equation}
that associates to each $p\in\mathcal{P}$ the push-forward probability measure of $\mu_\Omega$ under $f\circ\mathcal{E}_p\circ\xi$. Thus, $\mu_f(p)$ is the distribution of features of the ensemble of patterns generated by the randomized initial conditions at the parameter value $p\in\mathcal{P}$. If the feature space $\mathcal{Z}$ is a compact interval in $\R$, we can also define the \emph{feature mean} $E_f(p)\in\R$ via
\begin{equation}\label{e:featmean}
E_f\colon\quad \mathcal{P} \longrightarrow \R, \quad
p \longmapsto E_f(p) := E(\mu_f(p)) = \int_\mathcal{Z} z\,\rmd\mu_f(z;p) \in \mathcal{Z}.
\end{equation}
Our next result shows that $\mu_f$ and $E_f$ are continuous when we equip $\mathcal{P}$ and $\R$ with the standard Euclidean metric and the space $\Prob(\mathcal{Z})$ with the 2-Wasserstein metric $d_\mathrm{W}$ defined in (\ref{e:W2}).

\begin{Lemma}\label{l:4}
Assume that Hypothesis~\ref{h1} is met, then the map $\mu_f\colon (\mathcal{P},d_\mathrm{Eucl}) \to(\Prob(\mathcal{Z}),d_\mathrm{W})$ is continuous. Furthermore, if $\mathcal{Z}\subset\R$ is a compact interval, then the map $E_f\colon\mathcal{P}\to\mathcal{Z}$ is continuous.
\end{Lemma}

\begin{Proof}
We focus first on continuity of $\mu_f$. Using the results we quoted in our Digression, it suffices to show that $\mu_f(q)\rightharpoonup\mu_f(p)$ at $q\to p$ in $\mathcal{P}$. Using the definition
\[
F(\omega;p) := (f\circ\mathcal{E}_p\circ\xi)(\omega)
\]
we need to prove that for each fixed choice of $g\in C^0(\mathcal{Z},\R)$ we have
\begin{equation}\label{e:pfm}
\int_\mathcal{Z} g(z)\,\rmd\mu_f(z;q) = \int_\Omega g(F(\omega;q))\,\rmd\mu_\Omega \longrightarrow \int_\Omega g(F(\omega;p))\,\rmd\mu_\Omega = \int_\mathcal{Z} g(z)\,\rmd\mu_f(z;p) \quad\mbox{ as }\quad q \to p \mbox{ in } \mathcal{P}.
\end{equation}
Thus, fix $g\in C^0(\mathcal{Z},\R)$ and let $m:=\max_{z\in\mathcal{Z}}|g(z)|$. Pick $\epsilon>0$ and define $\mathcal{X}_\mathrm{s}:=\mathcal{X}\setminus\mathcal{X}_\mathrm{reg}$. Since $\mu_\mathcal{X}(p)(\mathcal{X}_\mathrm{s})=0$ by Hypothesis~\ref{h1}, there is a $\delta>0$ so that $\mu_\mathcal{X}(p)(U_\delta(\mathcal{X}_\mathrm{s}))\leq\frac{\epsilon}{4m}$, where $U_\delta(A)$ denotes the $\delta$-neighborhood of a set $A$ in $\mathcal{X}$. Hence, by definition, $\Omega_\delta:=(\mathcal{E}_p\circ\xi)^{-1}(U_\delta(\mathcal{X}_\mathrm{s}))$ satisfies $\mu_\Omega(\Omega_\delta)\leq\frac{\epsilon}{4m}$. We set $\Omega_\delta^\mathrm{c}:=\Omega\setminus\Omega_\delta$ and claim that for each fixed $\omega\in\Omega_\delta^\mathrm{c}$ the function $g(F(\omega;q))$ is continuous in $q$ for all $q\in\mathcal{P}$ near $p$: this claim is true since $(\mathcal{E}_q\circ\xi)(\omega)$ is continuous in $q$, the image lies in $\mathcal{X}_\mathrm{reg}$ for all $q$ near $p$ since $\omega\in\Omega_\delta^\mathrm{c}$, and $g\circ f$ is, by definition, continuous on $\mathcal{X}_\mathrm{reg}$. Lebesgue's dominated convergence theorem therefore implies that there is a $\tilde{\delta}>0$ so that
\[
\int_{\Omega_\delta^\mathrm{c}} \left|g(F(\omega;q)) - g(F(\omega;p))\right| \,\rmd\mu_\Omega < \frac{\epsilon}{2} \quad\mbox{ for all } q\in\mathcal{P} \mbox{ with } |q-p|<\tilde{\delta}.
\]
On $\Omega_\delta$, we have
\[
\int_{\Omega_\delta} \left|g(F(\omega;q)) - g(F(\omega;p))\right| \,\rmd\mu_\Omega \leq 2m \mu_\Omega(\Omega_\delta) < \frac{\epsilon}{2} \quad\mbox{ for all } q\in\mathcal{P}.
\]
We conclude that
\[
\int_{\Omega} \left|g(F(\omega;q)) - g(F(\omega;p))\right| \,\rmd\mu_\Omega < \epsilon \quad\mbox{ for all } q\in\mathcal{P} \mbox{ with } |q-p|<\tilde{\delta},
\]
which establishes weak convergence of $\mu_f(q)$. Finally, if $\mathcal{Z}$ is an interval in $\R$, the expression for $E_f(p)$ from (\ref{e:featmean}) coincides with the left-hand side of (\ref{e:pfm}) with $g(z)=z$, and continuity of $E_f(p)$ therefore follows from the arguments for weak convergence given above.
\end{Proof}

\subsubsection*{Empirical measures}

In practice, given a reaction-diffusion model (\ref{e:pde}) and a feature function $f$, we will not be able to compute the resulting pattern statistics analytically. Instead, we will approximate the pattern statistics numerically using empirical measures. In (\ref{e:xi}), we defined the function $\xi\colon\Omega\to\ell^2(\Z^2)$ that provided the randomization of initial conditions. The empirical measure will be based on a sequence of random variables with the same distribution as $\xi$.

\begin{Hypothesis}\label{h2}
Assume that $(\xi^{n})_{n\in\N}$ is a sequence of independent, identically distributed random variables $\xi^{n}\colon\Omega\to\ell^2(\Z^2)$ with $\xi^{n}\in L^2(\Omega,\ell^2(\Z^2))$ that satisfy $\xi^{n}_\#\mu_\Omega=\xi_\#\mu_\Omega$ for all $n$.
\end{Hypothesis}

We set $\xi^{n,\omega}:=\xi^{n}(\omega)$. For each fixed $\omega\in\Omega$ and each $N\geq1$, we then define the empirical measure
\[
\mu^{N,\omega}_{\ell^2} := \frac{1}{N} \sum_{n=1}^N \delta_{\xi^{n,\omega}}
\]
on the space $\ell^2(\Z^2)$. We can think of each empirical measure $\mu^{N,\omega}_{\ell^2}$ as arising from drawing $N$ independent samples from the space $\Omega$ with the measure $\mu_\Omega$ and constructing the resulting push-forward measure on $\ell^2(\Z^2)$ under the map $\xi$. For each fixed $N\in\N$ and $\omega\in\Omega$, we can then define the \emph{empirical pattern statistics} via 
\begin{eqnarray}\label{e:empfeatdistr}
\mu_f^{N,\omega} & \colon & \mathcal{P} \longrightarrow \Prob(\mathcal{Z}),\quad
p \longmapsto \mu_f^{N,\omega}(p) := (f\circ\mathcal{E}_p)_\# \mu^{N,\omega}_{\ell^2} =
\frac{1}{N} \sum_{n=1}^N \delta_{(f\circ\mathcal{E}_p)(\xi^{n,\omega})}.
\end{eqnarray}
As before, if the feature space $\mathcal{Z}$ is a compact interval in $\R$, the \emph{empirical feature mean} $E_f^{N,\omega}(p)\in\R$ is given by
\begin{equation}\label{e:empfeatmean}
E_f^{N,\omega}\colon\quad \mathcal{P} \longrightarrow \R, \quad
p \longmapsto E_f^{N,\omega}(p) := \int_\mathcal{Z} z\,\rmd\mu_f^{N,\omega}(z;p) =
\frac{1}{N} \sum_{n=1}^N (f\circ\mathcal{E}_p)(\xi^{n,\omega}) \in \mathcal{Z}.
\end{equation}
Our next result shows that for almost every $\omega\in\Omega$ these statistics converge to the full statistics as $N\to\infty$.

\begin{Lemma}\label{l:5}
Assume Hypotheses~\ref{h1}-\ref{h2} are met, then for each $p\in\mathcal{P}$ the sets
\[
\Omega_1:=\left\{\omega\colon d_\mathrm{W}\left(\mu_f^{N,\omega}(p),\mu_f(p)\right)\to0 \mbox{ as } N\to\infty\right\} \mbox{ and }
\Omega_2:=\left\{\omega\colon\left|E_f^{N,\omega}(p)-E_f(p)\right|\to0 \mbox{ as } N\to\infty\right\}
\]
satisfy $\mu_\Omega(\Omega_j)=1$ for $j=1,2$. Furthermore, \changed{for each fixed $p_0\in\mathcal{P}$ and $N\in\N$, the quantities $\mu_f^{N,\omega}(p)$ and $E_f^{N,\omega}(p)$ are continuous in $p$ at $p=p_0$ for almost every $\omega\in\Omega$.}
\end{Lemma}

\changed{Note that, for fixed $N$ and $\omega$, the empirical measure $\mu_f^{N,\omega}(p)$ may not be continuous in $p$ across open intervals in $p$ when the feature function $f$ is integer-valued.}

\begin{Proof}
Weak convergence of empirical measures for almost every $\omega$ was established in \citep[Theorem~3]{Varadarajan}, which yields convergence in the 2-Wasserstein distance by \citep[Corollary~6.13]{Villani}. As in the proof of Lemma~\ref{l:4}, weak convergence of $\mu_f^{N,\omega}(p)$ implies convergence of $E_f^{N,\omega}(p)$. It remains to prove continuity in $p$. We use again the notation $\mathcal{X}_\mathrm{s}:=\mathcal{X}\setminus\mathcal{X}_\mathrm{reg}$. Since the random variables $\xi^{n}$ have the same distribution as $\xi$, the set
\[
\Omega_3 := \left\{\omega\in\Omega\colon\mathcal{E}_p(\xi^{n,\omega})\in\mathcal{X}_\mathrm{s}\mbox{ for some }n\geq1\right\} 
= \bigcup_{n\geq1}\underbrace{\left\{\omega\in\Omega\colon\mathcal{E}_p(\xi^{n,\omega})\in\mathcal{X}_\mathrm{s}\right\}}_{\mbox{\small has measure zero by \ref{h1}}}
\]
has measure zero for each fixed $p\in\mathcal{P}$. For each $\omega\in\Omega_3$, we can now proceed as in Lemma~\ref{l:4} to prove continuity in $p$ for each $N\geq1$.
\end{Proof}

\subsubsection*{Objective functions}

Next, we compare the pattern statistics for different parameter values using the Wasserstein distance. We define the \emph{objective function} $G_f$ via
\begin{equation}\label{e:G}
G_f\colon\mathcal{P}\times\mathcal{P}\longrightarrow\R, \quad (p,q)\longmapsto
G_f(p,q) := d_\mathrm{W}(\mu_f(p),\mu_f(q))
\end{equation}
and the \emph{empirical objective function} $G_f^N$ via
\begin{equation}\label{e:Gemp}
G_f^N\colon\mathcal{P}\times\mathcal{P}\longrightarrow\R, \quad (p,q)\longmapsto
G_f^N(p,q) := d_\mathrm{W}\left(\mu_f^{N,\omega}(p), \mu_f^{N,\tilde{\omega}}(q)\right).
\end{equation}
In the empirical objective function, we use the same number of samples in both arguments as this allows us to use the discrete optimal-transport formulation (\ref{e:wot}) for the Wasserstein distance. In contrast, we will typically evaluate the two empirical measures in the argument at different elements $\omega,\tilde{\omega}\in\Omega$. If $\mathcal{Z}=[0,\mathfrak{m}]$, we will also use the objective functions
\begin{eqnarray}\label{e:Gmean}
G_{E_f} & \colon & \mathcal{P}\times\mathcal{P}\longrightarrow\R, \quad
(p,q)\longmapsto G_{f}E(p,q) := |E_f(p)-E_f(q)| \\ \nonumber
G_{E_f}^N & \colon & \mathcal{P}\times\mathcal{P}\longrightarrow\R, \quad
(p,q)\longmapsto G_{Ef}^N(p,q) := \frac{1}{N} \left|\sum_{n=1}^N (f\circ\mathcal{E}_p)(\xi^{n,\omega}) - \sum_{n=1}^N (f\circ\mathcal{E}_q)(\xi^{n,\tilde{\omega}}) \right|
\end{eqnarray}
that compare the feature means at different parameter values. Lemmas~\ref{l:4} and~\ref{l:5} show that the objective functions depend continuously on $p$ (almost surely in the case of empirical objective functions) and that $G_f^N$ and $G_{E_f}^N$ converge to $G_f$ and $G_{E_f}$, respectively, as $N\to\infty$ almost surely.

\subsubsection*{Bags and empirical pattern statistics}

For the case where $\mathcal{Z}=\Prob([0,\mathfrak{m}])$, we can also use an empirical pattern statistics that aggregates feature values of connected pattern components across samples. In this case, the empirical pattern statistics
\[
\mu_f^{N,\omega}(p) = \frac{1}{N}\sum_{n=1}^N\delta_{(f\circ\mathcal{E}_p)(\xi^{n,\omega})} \in \Prob(\mathcal{Z})
\]
has support on the elements $(f\circ\mathcal{E}_p)(\xi^{n,\omega})\in\mathcal{Z}=\Prob([0,\mathfrak{m}])$. We assume that these elements are finite sums of $\delta$-functions so that
\begin{equation}\label{e:fdelta}
(f\circ\mathcal{E}_p)(\xi^{n,\omega}) = \frac{1}{m^{n,\omega}_p} \sum_{j=1}^{m^{n,\omega}_p} \delta_{a^{n,\omega}_{j,p}} \in \mathcal{Z} = \Prob([0,\mathfrak{m}]), \qquad a^{n,\omega}_{j,p}\in[0,\mathfrak{m}]
\end{equation}
as is the case for the feature functions $f_\mathrm{AreaDistr}$ and $f_\mathrm{RoundDistr}$ that we defined earlier. We use the collection (the "bag") of feature values $a^{n,\omega}_{j,p}\in[0,\mathfrak{m}]$ across the probability measures $(f\circ\mathcal{E}_p)(\xi^{n,\omega})$ on $\Prob([0,\mathfrak{m}])$ to define the probability measure
\begin{equation}\label{e:Fbag}
\mu_{f,\mathrm{bag}}^{N,\omega}(p) := \frac{1}{N_\mathrm{bag}^\omega(p)} \sum_{n=1}^N \sum_{j=1}^{m^{n,\omega}_p} \delta_{a^{n,\omega}_{j,p}} \in \mathcal{Z} = \Prob([0,\mathfrak{m}]), \qquad
N_\mathrm{bag}^\omega(p) := \sum_{n=1}^N m^{n,\omega}_p
\end{equation}
on $[0,\mathfrak{m}]$. The empirical measure $\mu_{f,\mathrm{bag}}^{N,\omega}(p)$ can therefore be thought of as the distribution of the feature values of the individual connected components collected from the multiset of the $N$ pattern samples. The objective function
\[
G^N_{f,\mathrm{bag}}(p,q) = d_\mathrm{W}\left( \mu_{f,\mathrm{bag}}^{N,\omega}(p), \mu_{f,\mathrm{bag}}^{N,\tilde{\omega}}(q) \right)
\]
compares the resulting empirical measures using the Wasserstein distance on $\mathcal{Z}=\Prob([0,\mathfrak{m}])$.

Note that empirical pattern statistics $\mu_{f,\mathrm{bag}}^{N,\omega}(p)$ cannot, to our knowledge, be interpreted as sampling from an underlying distribution on $\Prob(\mathcal{Z})$. Furthermore, even though the sequence $\mu_{f,\mathrm{bag}}^{N,\omega}(p)$ will have convergent subsequences as $N\to\infty$ due to compactness of $\mathcal{Z}$, it is not clear whether the sequence itself converges and what the weak limit would be.

\subsubsection*{Illustration and intuition}

We illustrate the concepts we introduced above through a very simple example. Consider a reaction-diffusion system at a single parameter value and assume that the system generates only two distinct patterns $A$ and $B$. We assume that $A$ has three connected components each with area $a$, while $B$ is connected with area $1$. We encounter the set $A$ with probability $\rho\in[0,1]$ and the set $B$ with probability $1-\rho$. We set $I:=[0,\mathfrak{m}]$ and focus on the feature function $f_\mathrm{A}:=f_\mathrm{AreaDistr}$ that associates to a pattern $C$ the probability measure $f(C)=\frac{1}{\beta(C)}\sum_{j=1}^{\beta(C)}\delta_{|C_j|}$ in $\mathcal{Z}=\Prob(I)$ with atoms on the areas $|C_j|$ of the connected components $C_j$ of the pattern $C$, and its expectation $f_\mathrm{E_A}:=f_\mathrm{E_{AreaDistr}}=E(f_\mathrm{AreaDistr})=\frac{1}{\beta(C)}\sum_{j=1}^{\beta(C)}|C_j|$ in $I$.

We have $f_\mathrm{A}(A)=\frac{1}{3}\sum_{j=1}^{3}\delta_a=\delta_a$ and $f_\mathrm{A}(B)=\delta_1$ as well as $f_\mathrm{E_A}(A)=a$ and $f_\mathrm{E_A}(B)=1$. The pattern statistics $\mu_{f_\mathrm{A}}$ associated with the feature $f_\mathrm{A}$ is given by
\[
\mu_{f_\mathrm{A}} = \rho\delta_{f_\mathrm{A}(A)} + (1-\rho)\delta_{f_\mathrm{A}(B)}
= \rho\delta_{\delta_a}+(1-\rho)\delta_{\delta_1} \in \Prob(\Prob(I)).
\]
Furthermore, the feature mean of $f_\mathrm{E_A}$ is given by $E_{f_\mathrm{E_A}}=\rho f_\mathrm{E_A}(A)+(1-\rho)f_\mathrm{E_A}(B)=\rho a+(1-\rho)$, which provides the expectation of the average area of a single connected component in each individual pattern sample.

Next, we consider the empirical measure $\mu_{f_\mathrm{A},\mathrm{bag}}^{N}$ defined in (\ref{e:Fbag}) for the area distribution $f_\mathrm{A}$ from $N$ sampled patterns with $N\gg1$. We obtain
\[
\mu_{f_\mathrm{A},\mathrm{bag}}^{N} = \frac{1}{1+2\rho} (3\rho\delta_a + (1-\rho)\delta_1) \in \Prob(I), \qquad
E(\mu_{f_\mathrm{A},\mathrm{bag}}^{N}) = \frac{1}{1+2\rho}(3\rho a+(1-\rho))=\frac{1+\rho(3a-1)}{1+2\rho} \in I
\]
for the empirical measure and its expectation. The expectation $E(\mu_{f_\mathrm{A},\mathrm{bag}}^{N})$ gives the expected area of individual connected components in the multiset (the "bag") of all connected components across $N$ samples of the patterns $A$ and $B$. Note that $E(\mu_{f_\mathrm{A},\mathrm{bag}}^{N})$ and $E_{f_\mathrm{E_A}}$ are not equal, since the former corresponds to the expected area in the multiset of all connected components across all patterns, while the latter captures the expected area for the connected components in a typical pattern.

\subsubsection*{Spatial discretization}

We now discuss the numerical computation of pattern statistics, focusing first on the spatial discretization of the reaction-diffusion system (\ref{e:pde}) and then on the numerical approximation of patterns and sublevel sets using $\alpha$-shapes. To discretize (\ref{e:pde}) in space, we fix $K$ and approximate solutions of (\ref{e:pde}) via the orthogonal projection $\mathcal{Q}_K$ onto the closed subspace $H^2_K(D):=\{U=\sum_{k\in\Z^2,|k|\leq K} a_k e_k\}$ of $H^2(D)$ so that (\ref{e:pde}) becomes
\begin{equation}\label{n:pde}
U_t = \mathcal{D} \Delta U + \mathcal{Q}_K \mathcal{N}(U,p), \quad U\in H^2_K(D).
\end{equation}
The solutions of (\ref{n:pde}) are of the form $U_K(t)=\sum_{k\in\Z^2,|k|\leq K} a_k(t) e_k\in H^2_K(D)\subset H^2(D)$, and we write $\Phi_T^K(U_K(0),p):=U_K(T)$. We can use the randomization of initial conditions we introduced before projected by $\mathcal{Q}_K$ onto the set of Fourier coefficients corresponding to wavenumbers $k$ with $|k|\leq K$. The remaining part of the framework remains unchanged since $\Phi_T^K(U,p)$ still lies in $H^2(D)$. For each fixed $U_0\in H^2(D)$, we have $\Phi_T^K(\mathcal{Q}_K U_0,p)\to\Phi_T(U_0,p)$ in $H^2(D)$ as $K\to\infty$ (see \citep{Gottlieb, Hald}). We note that choosing our randomized initial data from $H^5(D)$ instead if $H^2(D)$ would also ensure regularity in $C^2(D)$ \changed{\cite[\S3.5]{Henry}}. Using these convergence properties, we can proceed analogously to Lemma~\ref{l:4} to show that the pattern statistics associated with (\ref{n:pde}) converges weakly, and hence also in the Wasserstein distance, to the statistics associated with (\ref{e:pde}) as $K\to\infty$. In particular, discretization of the reaction-diffusion system allows us to faithfully approximate the pattern statistics of the full system.

\subsubsection*{Computation of feature functions via $\alpha$-shapes}

\begin{figure}
\centering
\includegraphics[scale=1.35]{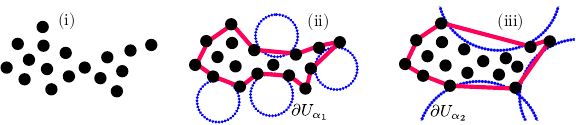}
\caption{We illustrate the definition of $\alpha$-shapes and their dependence on the radius $\alpha$. Panel~(i) shows the original data set. The associated $\alpha$-shapes for radii $\alpha_1$ and $\alpha_2$ with $\alpha_1<\alpha_2$ are shown as polygons in panels~(ii) and~(iii), respectively. The circles $\partial U_{\alpha_{1,2}}$ determine which data points are connected by edges.}
\label{f:alpha}
\end{figure}

Next, we discuss the numerical computation of sublevel sets. We choose $M\geq1$ and define the finite lattice $D_M$ consisting of $M^2$ equally spaced points in the domain $D=(\R/2\pi\Z)^2$. For $u\in\mathcal{X}_\mathrm{reg}$, the sublevel set $A:=\{x\in D\colon u(x)\leq c\}$ is a 2-manifold with boundary of class $C^2$. The pattern belonging to $u$ on $D_M$ is then given by the discrete set $A_M:=A\cap D_M=\{x\in D_M\colon u(x)\leq c\}$ of points $x$ on the lattice $D_M$ for which $u(x)\leq c$. We will use $\alpha$-shapes to approximate the number of connected components of $A$, the length $\lambda(\partial A)$ of its boundary, and its area $\mu_\mathrm{Leb}(A)$.

Alpha-shapes are defined as follows (see also Figure~\ref{f:alpha} for an illustration). For fixed $\alpha>0$, the $\alpha$-shape $S_\alpha(A_M)$ of $A_M$ is a disjoint union of polygons \changed{with vertices in $A_M$} whose edges are defined as follows \citep{Edelsbrunner}: two elements $x_i, x_j\in A_M$ form an edge in $S_\alpha(A_M)$ if and only if there is an open ball $U_\alpha$ of radius $\alpha$ so that $x_i,x_j\in \partial U_\alpha$ and $A_M\cap U_\alpha=\emptyset$. 

As shown in \changed{\citep[\S{}II and IV.B]{Edelsbrunner}, there is a unique closed set $F\subset D$ so that $\partial F=S_\alpha(A_M)$ and $A_M\subset F$. Furthermore, $F$ has open interior. We set $\mathring{S}_\alpha(A_M):=F$ and refer to it as interior face of $S_\alpha(A_M)$.} In preparation for the next lemma, we say that a set $A$ satisfies the $r$-rolling condition if for each $x\in A$ there is an open ball $U_r$ of radius $r$ so that $x\in\partial U_\alpha$ and $A\cap U_r=\emptyset$. It was shown in \citep[Theorem~1]{Walther} that if $A$ is a compact 2-manifold of class $C^2$, then there is an $r>0$ so that $A$ and $A^\mathrm{c}$ satisfy the $r$-rolling condition: we denote this radius by $r(A)$.

\begin{Lemma}\label{l:alpha}
Let $A$ be a 2-manifold with $C^2$ boundary in $D$, and fix $\alpha\in(0,r(A))$, then there are constants $C_0,M_0$ that depend only on $(\alpha,r(A))$ so that the following is true for all $M\geq M_0$. The set $A$ and the interior face $\mathring{S}_\alpha(A_M)$ have the same number $\beta(A)$ of connected components, denoted by $A_j$ and $\mathring{S}_\alpha^j(A_M)$, respectively, and these can be labeled so that $A_j\cap D_M\subset\mathring{S}_\alpha^j(A_M)$ for $j=1,\ldots,\beta(A)$. Furthermore, we have
\begin{equation}\label{e:alpha}
\left| \mu_\mathrm{Leb}(A)-\mu_\mathrm{Leb}(\mathring{S}_\alpha(A_M)) \right| +
\max_{j=1,\ldots,\beta(A)} \left| \lambda(\partial A_j)-\lambda(S_\alpha^j(A_M)) \right| \leq \frac{C_0}{M}
\end{equation}
for each $M\geq M_0$.
\end{Lemma}

\begin{Proof}
The lemma follows from the results in \citep{AriasCastro}. While that paper focused primarily on the case of independent samples from the uniform distribution in $A$, many of the results are for the deterministic case, and we will now show how they can be used to prove our claims. We will be very brief and refer to \citep{AriasCastro} for details and notation. First, we will show that the set $H^\mathrm{c}_{ij,t}\cap G_{ij}$ considered in \citep[Proposition~5]{AriasCastro} is not just of small measure but is indeed empty in our situation provided $t\geq 8\pi C_1/M$, where $C_1=C_1(\alpha,r(A))$ denote the constant defined in \citep[Lemma~5]{AriasCastro}. Assume that $x_i,x_j$ form an edge in $S_\alpha(A_M)$ so that $x_i,x_j\in U_\alpha(z)$ for $z\notin S$, where $U_\alpha(z)$ denotes the open ball of radius $\alpha$ centered at $z$. If $\alpha-d(z,S)\geq 8\pi/M$, then we can follow the arguments in \citep[Lemma~5]{AriasCastro} to show that $U_\alpha(z)\cap A_M\neq\emptyset$, since each square of length $4\pi/M$ inside $A$ necessarily contains an element in $A_M$. In particular, if $t\geq 8\pi C_1/M$ and $0<d(z,S)<\alpha-t/C_1$, then we have $8\pi/M\leq t/C_1<\alpha-d(z,S)$ and therefore $U_\alpha(z)\cap A_M\neq\emptyset$, which shows that the set $H^\mathrm{c}_{ij,t}\cap G_{ij}$ considered in \citep[Proposition~5]{AriasCastro} is empty. The arguments in \cite[\S4]{AriasCastro} now show that $|\lambda(\partial A)-\lambda(S_\alpha(A_M))|\leq C_0 t$ for all $t$ with $t\geq 8\pi C_1/M$. Choosing $t=8\pi C_1/M$ completes the proof of the perimeter estimate. Using the fact that \citep[Propositions~1 and~5]{AriasCastro} are true in our situation provided $t\geq 8\pi C_1/M$, the statements in \citep[Propositions~5 and~6]{AriasCastro} also hold and show that the distance between $\partial A$ and $S_\alpha(A_M)$ is bounded by $C_2/M$, where $C_2$ depends only on $(\alpha,r(A))$. This fact can now be used to establish the estimate for the difference of the areas of $A$ and $\mathring{S}_\alpha(A_M)$ and to show the statements about their connected components.
\end{Proof}

In summary, the pattern statistics of (\ref{e:pde}) is accurately approximated under spatial discretization provided $K\gg1$ is large enough. Similarly, the features of each fixed sublevel set are approximated at order $\rmO(1/M)$ when we compute them using the $\alpha$-shapes on the lattice $D_M$ with $M^2$ points: we remark that we do not have uniformity in $A=u^{-1}((-\infty,c])$, since $u$ may be arbitrarily close to $\mathcal{X}\setminus\mathcal{X}_\mathrm{reg}$ where $c$ is no longer a regular value.


\section{Tracing bifurcations using pattern statistics} \label{s:3}

We build on the framework outlined in \S\ref{s:2} to trace out curves in the two-dimensional parameter space $\mathcal{P}\subset\R^2$ that separate regions in parameter space with different prevailing patterns. If $f$ is a feature function that can distinguish the patterns we are interested in, we will argue in this section that the objective function $G_f(p,q)=d_\mathrm{W}(\mu_f(p),\mu_f(q))$, which measures the difference between the associated pattern statistics $\mu_f$ at different parameter values, can be used to characterize and compute transition curves using predictor-correct continuation.

\begin{figure}
\centering
\includegraphics[scale=1.1]{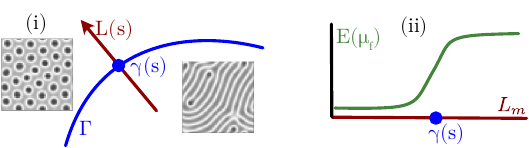}
\caption{Panel~(i) illustrates the curve $\Gamma$ that separates parameter regions where spots and stripes prevail. The feature function $f_\mathrm{Conn}$ assigns to each pattern the number of its connected components: as indicated in panel~(ii), the expectation $E(\mu_f)$ of its pattern statistics should therefore change from low to high values as we cross from the stripe into the spot region.}
\label{f:trans}
\end{figure}

\subsubsection*{Bifurcation functions}

We illustrate the concepts by focusing first on the feature function $f=f_\mathrm{Conn}$ that counts the connected components in a given pattern. This feature function will distinguish spots and stripes. Assume that $\gamma\colon\R\to\mathcal{P}\in C^1$ is a curve so that its trace \changed{$\Gamma=\{\gamma(s)\colon s\in\R\}\subset\mathcal{P}$} separates regions of parameter space where each of these patterns prevails; see Figure~\ref{f:trans}(i) for an illustration. \changed{For each fixed $s\in\R$, we denote by $\mathfrak{n}(s)$ the unit vector normal to the tangent vector $\gamma^\prime(s)$ at a point $\gamma(s)\in\Gamma$ and define $L(s):=\gamma(s)+\R\mathfrak{n}(s)$ to be the line segment perpendicular to the curve $\Gamma$ at $\gamma(s)$.} As indicated in Figure~\ref{f:trans}(ii), the expectation $E(\mu_f(p))$ of the pattern statistics $\mu_f(p)$ will transition between two distinct states as the parameter $p$ crosses $\Gamma$, and the function $E(\mu_f(p))$ changes most rapidly at $p=\gamma(s)$ as $p$ varies in $L(s)$. Alternatively, we can choose a small offset $0<h\ll1$ and consider the rate of change of $E(\mu_f(p))$ measured by the slope
\[
\frac{1}{2h} \left| E(\mu_f(p+h\mathfrak{n}(s)))-E(\mu_f(p-h\mathfrak{n}(s))) \right|
\]
of the secant of the graph of $E(\mu_f(p))$: this slope will be largest at $p=\gamma(s)$, reflecting the fact that $E(\mu_f(p))$ changes most rapidly near $\gamma(s)$. Thus, we can compute $\gamma(s)$ via
\[
\gamma(s) = \argmax_{p\in L(s)} \frac{1}{2h} \left| E(\mu_f(p+h\mathfrak{n}(s)))-E(\mu_f(p-h\mathfrak{n}(s))) \right|.
\]
In general, we are interested in the pattern statistics $\mu_f(p)$, and the discussion above shows that we need to identify points $p$ where $\mu_f(p)$ changes most rapidly. To formalize this, we use the objective function
\[
G_f\colon \mathcal{P}\times\mathcal{P}\longrightarrow\R^+, \quad
(p,q)\longmapsto G_f(p,q) = d_\mathrm{W}(\mu_f(p),\mu_f(q))
\]
defined in (\ref{e:G}), which measures the 2-Wasserstein distance between the pattern statistics evaluated at the two parameter values $p$ and $q$ (in practice, we would use the empirical objective function $G_f^N$ defined in (\ref{e:Gemp})). Given a small offset $0<h\ll1$, we then expect that the \emph{bifurcation function}
\begin{equation}\label{e:g}
g(p; h) := \frac{1}{2h} G_f(p-h\mathfrak{n}(s),p+h\mathfrak{n}(s)), \qquad p\in L(s),
\end{equation}
which measures the rate of change of $\mu_f(p)$ at $p$, is maximized at $p=\gamma(s)$. In particular, given the line segment $L(s)$, we can find the intersection $\gamma(s)=\Gamma \cap L(s)$ by maximizing $g(p;h)$ along $L(s)$ so that
\begin{equation}\label{e:argmaxg}
\gamma(s) = \argmax_{p\in L(s)} g(p;h).
\end{equation}
We now describe how we can use the arg-max formulation (\ref{e:argmaxg}) to approximate $\Gamma$ numerically.

\subsubsection*{Predictor-corrector continuation of pattern statistics}

We outline our approach of using the bifurcation function $g(p)$ defined in (\ref{e:g}) to trace out a curve in parameter space $\mathcal{P}$ that separates regions with different prevailing patterns. We will postpone a discussion of how the function $g(p)$ can be evaluated to the next section and instead focus here on predictor-corrector continuation, assuming that we can evaluate $g(p)$ numerically. The main issues we need to tackle are (i) that $g(p)$ will be continuous but not necessarily differentiable and (ii) that the definition of $g$ as a distance requires the computation of the pattern statistics at two distinct points.

We will assume that the feature function $f$ can distinguish the patterns we are interested in. Our algorithm depends on the choice of the stepsize $s$ along the curve we want to compute and the offset $h$ in the computation of $g$. We assume that $h$ and $s$ have been chosen so that $0<h,s\ll1$.

\paragraph{Initialization}

\changed{To initialize the algorithm, we need to find an inial point $p_0\in\mathcal{P}$ so that $p_0$ lies on the transition curve $\Gamma$ we want to compute. We accomplish this by (1) selecting a line in the parameter space $\mathcal{P}$, (2) evaluating the selected feature function at equidistantly spaced points on the line (we choose the same distance $h$ as in the predictor-corrector algorithm), and (3) compute the Wasserstein distance of the resulting pattern statistics at consecutive points on the line. Each point $p_0$ at which the Wasserstein distance jumps is a potential bifurcation point from which we can start predictor-corrector continuation. To do so, we pick a second point $p_1\in\mathcal{P}$ by searching on the circle $|p_1-p_0|=s$ for a jump in Wasserstein distance and then apply the predictor-corrector steps outlined next.}

\begin{figure}
\centering
\includegraphics{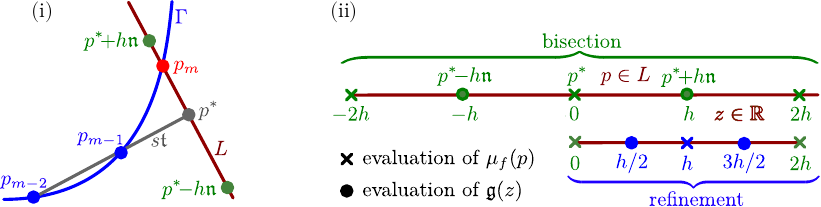}
\caption{Panel~(i) outlines the geometry for the predictor-corrector step, where we omitted the subscript $m$ in the notation. Panel~(ii) indicates at which points we evaluate the bifurcation function $\mathfrak{g}(z)$ (labeled by disks) and how this translates into evaluations of the pattern statistics $\mu_f(p)$ (labeled by crosses). The points in green correspond to the initial bisection in step~1., while the points in blue correspond to the refinement in step~2.\ for the case $\sigma=1$.}
\label{f:cont}
\end{figure}

\paragraph{Predictor-corrector steps}

We assume that we have successively computed the distinct points $p_0,\ldots,p_{m-1}\in\mathcal{P}$ on the curve $\Gamma$ for some $m\geq2$ and now outline how we determine the next point $p_m$ on the curve $\Gamma$. We define
\[
\mathfrak{t}_m := \frac{p_{m-1}-p_{m-2}}{|p_{m-1}-p_{m-2}|}, \quad
\mathfrak{n}_m := \mathfrak{t}_m^\perp, \quad
p_m^* := p_{m-1} + s \mathfrak{t}_m, \quad
L_m := \left\{ p\in\mathcal{P}\colon p=p_m^*+z\mathfrak{n}_m,\, z\in\R \right\}
\]
and refer to Figure~\ref{f:cont}(i) for an illustration of the underlying geometry. Our goal is to find $p_m$ as the solution to
\begin{equation}\label{e:objp}
p_m := \argmax_{p\in L_m} g_m(p;h), \qquad
g_m(p;h) := \frac{1}{2h} G_f(p-h\mathfrak{n}_m,p+h\mathfrak{n}_m).
\end{equation}
Using the linear parametrization $P_m(z):=p_m^*+z\mathfrak{n}_m$ of $L_m$ by $z\in\R$, we rewrite (\ref{e:objp}) equivalently as
\begin{equation}\label{e:objz}
p_m := P_m\left(\argmax_{z\in\R} \mathfrak{g}_m(z;h)\right), \qquad
\mathfrak{g}_m(z;h) := \frac{1}{2h} G_f(p_m^*+(z-h)\mathfrak{n}_m, p_m^*+(z+h)\mathfrak{n}_m).
\end{equation}
We showed in Lemmas~\ref{l:4} and~\ref{l:5} that $\mathfrak{g}_m(z;h)$ is continuous in $z$. Even though we do not know whether this function is differentiable, we will assume in step~3.\ below that $\mathfrak{g}_m(z;h)$ has a limit $\mathfrak{g}_m(z)$ as $h\to0$ and use this limit to motivate the use of quadratic interpolation to find the maximum of $\mathfrak{g}_m(z;h)$. We approximate the solution of (\ref{e:objz}) in the following three steps and refer to Figure~\ref{f:cont} for an illustration of these steps:
\begin{compactenum}[1.]
\item\textbf{Bisection:} Evaluate $g^+:=\mathfrak{g}_m(h;h)$ and $g^-:=\mathfrak{g}_m(-h;h)$.
\item\textbf{Refinement:} Set $\sigma:=\mathop{\mathrm{sign}}(g^+-g^-)$, and evaluate $g_{\sigma/2}:=\mathfrak{g}_m(\frac{h\sigma}{2};\frac{h}{2})$ and $g_{3\sigma/2}:=\mathfrak{g}_m(\frac{3h\sigma}{2};\frac{h}{2})$.
\item\textbf{Interpolation:} Determine the quadratic function $\mathfrak{g}_m^{(2)}(z)=a_2z^2+a_1z+a_0$ that passes through the points $g_{-\sigma}$, $g_{\sigma/2}$, and $g_{3\sigma/2}$ defined in step~2.\ for $z=-\sigma h$, $z=\frac{\sigma h}{2}$, and $z=\frac{3\sigma h}{2}$, respectively. If $a_2<0$ and $z_m^*=\frac{-a_1}{2a_2}\in(\min\{-\sigma h,\frac{3\sigma h}{2}\},\max\{-\sigma h, \frac{3\sigma h}{2}\})$, then we accept $p_{m+1}=p_m^*+z_m^*\mathfrak{n}_m$.
\end{compactenum}
We emphasize that step~1 requires the evaluation of the pattern statistics $\mu_f(p)$ at the points $p=p_m^*+jh\mathfrak{n}_m$ for $j=0,\pm1$. To complete step~2, assuming for simplicity that $\sigma=1$, we need to evaluate $\mu_f(p)$ only at the additional point $p=p_m^*+2h\mathfrak{n}_m$. We note that it is possible to replace the quadratic interpolation with additional bisection refinements: we found that quadratic interpolation produces better results in our numerical case studies.

We refer to Figure~\ref{fig:illustration} for a practical implementation of the algorithm described above: the colored disks in the center panel correspond to the parameter values at which the pattern statistics was computed for predictor-corrector step as outlined above.


\section{Assessing dependence on algorithmic parameters} \label{s:35}

We now outline the parameters that enter the algorithm and discuss how they affect accuracy and computational efficiency. The implementation of the continuation algorithm described above requires the following choices:
\begin{compactenum}[1.]
\item\textbf{Feature function:} We choose a feature function $f$ that can differentiate between the patterns we want to distinguish.
\item\textbf{Spatial discretization:} We choose the number $K$ of Fourier modes so that we can resolve the nonlinearity and the expected patterns at the expected wavelength. Alternatively, and this is how our numerical computations were conducted, we can use finite differences to solve the PDE model for a sufficiently small spatial stepsize that resolves the patterns we are interested in and use the resulting grid also for $\alpha$-shapes.
\item\textbf{Initial data:} \changed{We choose two fixed functions $U_\mathrm{b}(x)$ and $U_\mathrm{r}(x)$ to construct the initial data in (\ref{e:iota}). The deterministic part $U_\mathrm{b}$ is selected to ensure that we reach the patterns we are interested in: for domain-filling patterns, $U_\mathrm{b}$ is typically an unstable homogeneous rest state. The function $U_\mathrm{r}$ that will be used for the randomized part is given by $U_\mathrm{r}(x)=\sum_{|k|\leq K}a_ke_k(x)$ for a fixed nonzero choice of coefficients $(a_k)_{|k|\leq K}$ (we set $U_\mathrm{r}=0$ for deterministic initial data).}
\item\textbf{Randomization:} \changed{We select $N$ samples of the random variables $(b_k(\omega))_{|k|\leq K}$ from a uniform distribution and form the ensemble of $N$ randomized functions $U_\mathrm{r}^\omega(x):=\sum_{|k|\leq K}a_kb_k(\omega)e_k(x)$. The resulting $N$ functions $U_0(x)=U_\mathrm{b}(x)+U_\mathrm{r}^\omega(x)$ are then used as initial data in the numerical solver.} The number $N$ of samples that are used to calculate the empirical measure $\mu_f^{N,\omega}$ (or the empirical feature mean $E_f^{N,\omega}$ if applicable) can be adjusted using, for instance, a small-sample paired t-test (which tests the null hypothesis that the mean of the difference of feature samples is zero) to ensure that there is a statistically significant difference between the two empirical measures in the argument of the bifurcation function $\mathfrak{g}$.
\item\textbf{Integration time:} The integration time $T>0$ is chosen so that we reach the relevant pattern regime from the initial data $U_\mathrm{b}+U_\mathrm{r}^\omega$ within the time interval $[0,T]$. It is possible to adapt $T$ during continuation, for instance by choosing shorter or longer values and comparing the resulting feature values.
\item\textbf{Sublevel sets:} We evaluate the feature function on the sublevel sets $U_j^{-1}((-\infty,c])$ of the $j$th component of the solution $U$ to (\ref{e:pde}). We choose the index $1\leq j\leq d$ and the threshold $c\in\R$ so that the corresponding sublevel sets best reflect the patterns we are interested in.
\item\textbf{Alpha-shapes:} The evaluation of the feature functions we consider require the computation of the $\alpha$-shapes of the sublevel sets. We need to pick the radius $\alpha$ of the $\alpha$-shape and the number $M^2$ of lattice points on which we evaluate the solution to approximate the sublevel set. We usually choose $M:=K$ and set $\alpha=10/M$. We note that we can adapt $M$ and $\alpha$ by comparing the resulting pattern statistics using the Wasserstein metric to ensure that they do not change upon increasing $M$ or varying $\alpha$.
\item\textbf{Predictor-corrector steps:} We need to choose the arclength stepsize $s$ and the parameter offset $h$. We normally pick $h:=s$ and note that the stepsize $s$ can be adapted based on the successive changes of the angle of the secants, which are indicative of the curvature of the curve $\Gamma$.
\end{compactenum}

\begin{figure}
\centering
\includegraphics[width=0.9\textwidth]{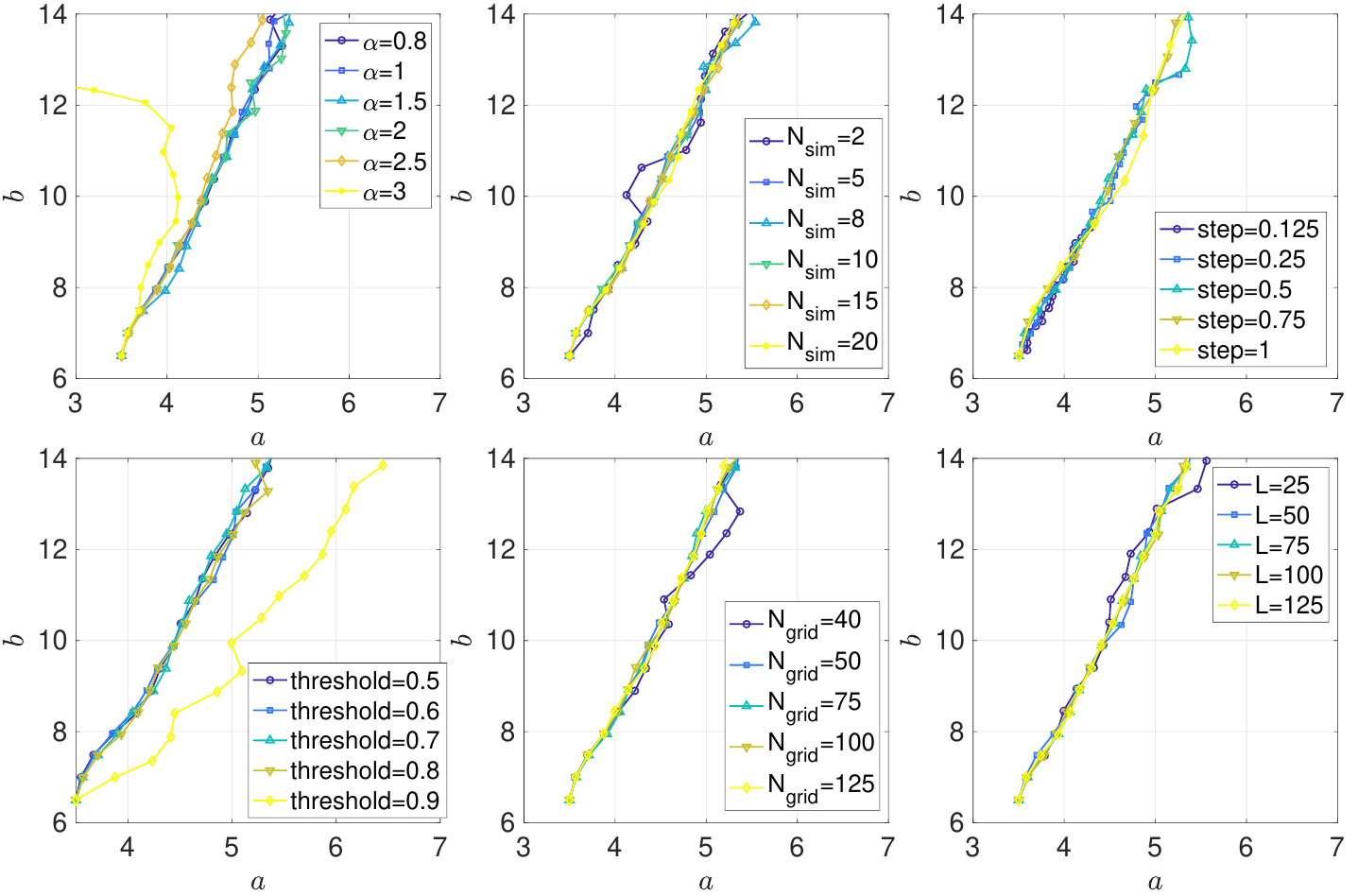}
\caption{\changed{Shown are the results of tracing out the spot-stripe coexistence curve for the Brusselator when changing the $\alpha$-shape parameter $\alpha$, the ensemble size $N=N_\mathrm{sim}$, the stepsize $s$, threshold parameter $c$, grid size $K=N_\mathrm{grid}$, and domain size $L$. The parameters are otherwise fixed at $\alpha=1$, $N=10$, $s=0.5$, $c=0.7$, $K=50$, and $L=50$.}}
\label{f:accuracy}
\end{figure}

\changed{Our algorithm is robust with respect to these choices, and we never had to adjust them during continuation. Generally, increasing the number $K$ of Fourier modes (or mesh points when using finite differences) for the PDE solver, the number $M$ of lattice points on which we evaluate patterns and their features, and the ensemble size $N$ of randomized initial data  will provide smoother and more accurate continuation curves. The algorithm is also robust with the respect to the choice of integration time $T$ (as long as $T$ is large enough so that the patterns of interest can be reached from randomized data within time $T$), the threshold $c$ (as long as the threshold is such that its sublevel sets contain the relevant information needed for the feature functions), and the parameter $\alpha$ that is used for the computation of the $\alpha$-shapes when evaluating feature functions. The default value we use for $\alpha$ is $\alpha=10/M$. Lemma~\ref{l:alpha} states that, for each $\alpha$ that does not exceed the rolling-ball condition for any sublevel set we want to capture, choosing $M$ large will accurately capture the relevant geometric features such as area and perimeter: choosing $\alpha$ too small for a given value of $M$ will result in very rough $\alpha$-shape boundaries so that the perimeter cannot be computed accurately. In Figure~\ref{f:accuracy}, we show the effect of different choices of the $\alpha$-shape parameter $\alpha$, ensemble size $N$, stepsize $s$, threshold parameter $c$, grid size $K$, and domain size $L$ on the coexistence curve between spots and stripes for the Brusselator model (this transition curve will be discussed in more detail in the next section).}

\changed{We will further validate the numerical accuracy of the proposed algorithm by comparing the results to numerical, theoretical, and mean-field approaches in Figures~\ref{f:pearson}, \ref{f:validation}, \ref{f:convection}, \ref{f:comparison}, \ref{f:snaking}, and~\ref{f:abm} below. In each of these cases, our algorithm reproduces known results accurately.}

\changed{The proposed continuation algorithm is computationally efficient. Since the computation of $\alpha$-shapes and Wasserstein distances is very fast, the main computational bottleneck are the direct PDE simulations. The time needed to compute ensembles of $N$ solutions can be reduced through the use of parallel processors and, as demonstrated in Figure~\ref{f:accuracy}, the algorithm produces accurate results even for solutions that are not highly resolved due to the statistical averaging effect across ensembles.}


\section{Results} \label{s:4}

In this section, we summarize the results of computations of bifurcation and transition curves for several common PDE models that utilize the predictor-corrector approach introduced in \S\ref{s:3} to trace out curves based on feature functions and pattern statistics. Details on the model systems to which we apply our approach can be found in the appendix. Our code is publicly available \citep{GithubRepo}, and we will therefore not discuss the numerical schemes we used in detail.


\begin{figure}[t]
\centering
\includegraphics[scale=0.9]{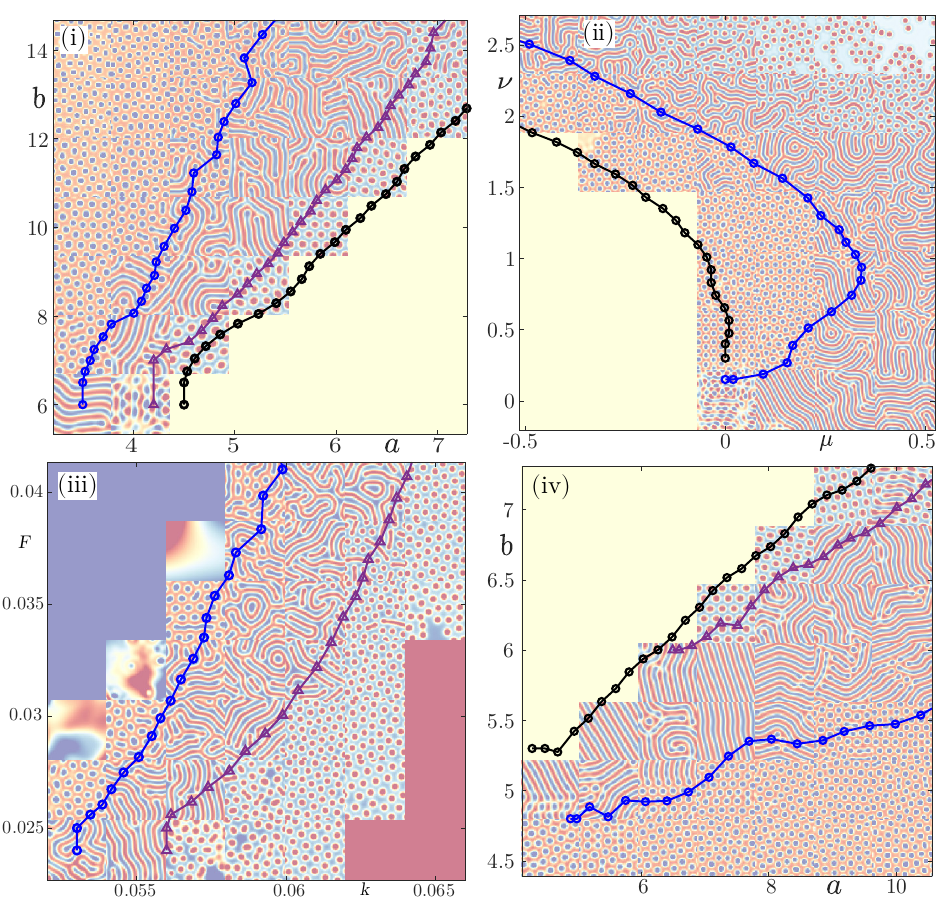}
\caption{Shown are curves that separate transitions from homogeneous states \changed{(which correspond to single-color tiles)} to spots or stripes (black circles), \changed{red spots} to stripes (purple triangles), and stripes to \changed{blue spots} (blue circles) for the (i)~Brusselator, (ii)~Swift--Hohenberg, (iii)~Gray--Scott, and (iv)~Schnakenberg models. The tiles show the results of direct numerical simulations for parameters set at the center of each tile to illustrate and validate the computed transition curves. \changed{We remark that stable stripes can bifurcate in the Swift--Hohenberg equation near $\nu=0$ due to the additional reflection symmetry $u\mapsto-u$; see \cite[\S5.4.1]{Hoyle}. Since we use a square domain, the different stability properties of rolls under square and hexagonal symmetries compete: on square domains, stable rolls can bifurcate \cite[\S5.3]{Hoyle} as seen in panel~(iv); the codimension-two point at $(a,b)\approx(6,6)$ in panel~(iv) corresponds to the transition from stable rolls to stable squares that occurs along the diagonal in \cite[Figure~4.10]{Hoyle}.}}
\label{f:coexistence}
\end{figure}

\begin{figure}
\centering
\includegraphics[width=0.9\textwidth]{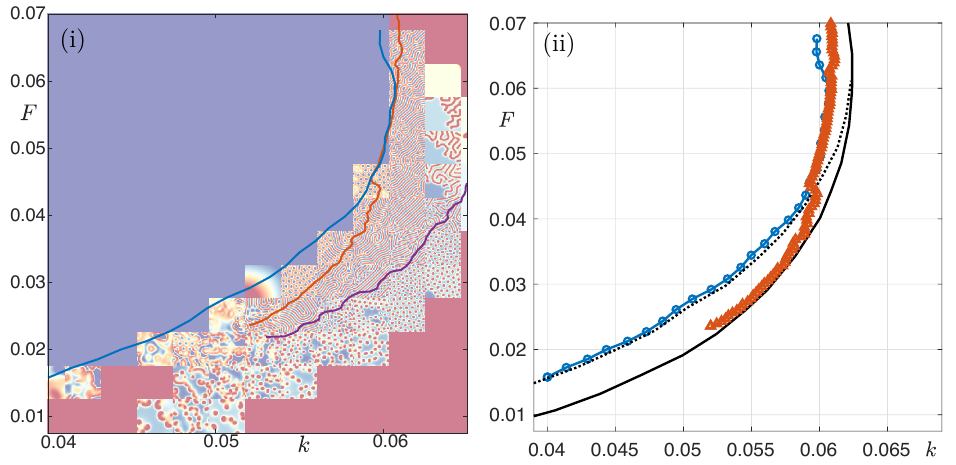}
\caption{\changed{Panel~(i) shows a larger region on the $(F,k)$-parameter space of the Gray--Scott model with an instability curve of the blue rest state in blue, a transition curve from blue spots to stripes in red, and a transition curve from stripes to red spots in purple, which were computed using the continuation approach proposed here. Panel~(ii) contains a comparison with the classification curves shown in \cite[Figure~3]{Pearson}, where the solid and dotted curves  correspond, respectively, to fold and subcritical Hopf bifurcations of the blue state; the red state is stable everywhere.}}
\label{f:pearson}
\end{figure}

\subsubsection*{Boundaries of coexistence regions}

First, we consider Turing bifurcations, along which a homogeneous rest state destabilizes and patterned states emerge, and transition curves between spot and stripe patterns. For both scenarios, we use randomized perturbations of the homogeneous rest state to create an ensemble of initial conditions. The sublevel sets of the resulting solutions are computed using either $U_1^{-1}([c(0.7),\infty))$ or $U_1^{-1}((-\infty,c(0.3)])$ with $c(s):=s\max(U_1)+(1-s)\min(U_1)$ depending on the type of spots we are interested in. To compute the pattern statistics, we use the bagged empirical probability measure $\mu_{f_\mathrm{RoundDistr},\mathrm{bag}}^{N}$, defined in (\ref{e:Fbag}), for the roundness-score feature function $f_\mathrm{RoundDistr}$ of $\alpha$-shapes.

\begin{figure}
\centering
\includegraphics[width=0.9\textwidth]{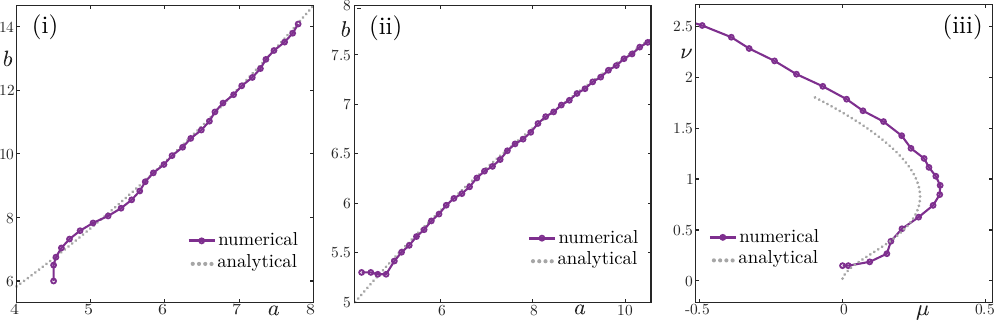}
\caption{In panels~(i) and~(ii), we compare the analytical Turing bifurcation curves, which correspond to transitions from homogeneous states to spots or stripes, to the curves found using our numerical algorithm for the Brusselator (left) and Schnakenberg (center) models. In panel~(iii), we compare the numerically computed curve that delineates transitions from spots to stripes in the Swift--Hohenberg model to the Maxwell curve along which spots and stripes have the same PDE energy and Lagrangian, where we computed the Maxwell curve using AUTO \cite{auto07p}.}
\label{f:validation}
\end{figure}

Figure~\ref{f:coexistence} shows our results for the curves that correspond to Turing bifurcations and transitions from spots to stripes in the Brusselator, Swift--Hohenberg, Gray--Scott, and Schnakenberg models. The figure also contains the results of typical direct numerical simulations as tiles to facilitate comparison of direct simulations with the transition curves computed using continuation. \changed{Figure~\ref{f:pearson} contains a comparison of our continuation curves with those in \cite[Figure~3]{Pearson}.} Figure~\ref{f:validation} shows comparisons of the analytical expressions of Turing bifurcation curves for the Brusselator and Schnakenberg models with the curves we computed numerically as well as a comparison of the transition curve between spots and stripes in the Swift--Hohenberg equation with the associated Maxwell curve. The Swift--Hohenberg equation is variational, and the Maxwell curve corresponds to parameter values where spots and stripes, for the wave number that destabilizes first at the Turing bifurcation, have the same PDE energy and Lagrangian, and should therefore co-exist along this curve. The results illustrated in Figures~\ref{f:coexistence} and~\ref{f:validation} show very good agreement between the curves computed using our approach and comparisons with direction numerical simulations and analytical bifurcation curves. We note that the discrepancy between the Maxwell curve and the transition curve for the Swift--Hohenberg equation is likely due to the fact that the wave number selected by spots and stripes deviates from the wave number selected at the Turing bifurcation curve $\mu=0$.

\begin{figure}
\centering
\includegraphics[width=0.9\textwidth]{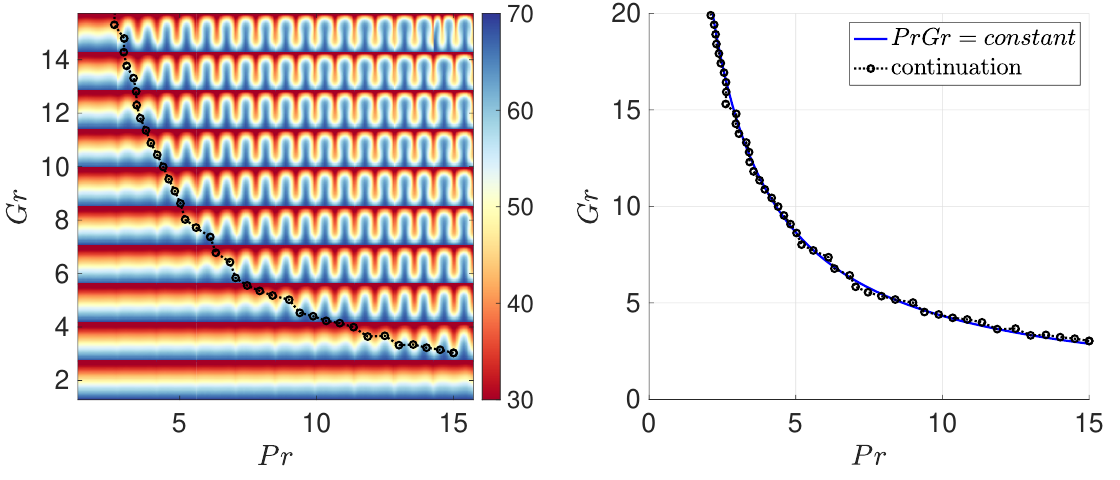}
\caption{Shown is the transition curve between conducting and convecting patterns in the $(\mathrm{Pr},\mathrm{Gr})$ parameter space and its validation again the theoretically predicted curve for the Rayleigh--B\'enard model.}
\label{f:convection}
\end{figure}

\changed{Figure~\ref{f:convection} shows an application to the two-dimensional incompressible Rayleigh--B\'enard fluid flow. Depending on the Prandtl and Grashof numbers (Pr and Gr, respectively), this fluid flow accommodates a temperature difference across the top and bottom of its surrounding container through heat conduction with a linear temperature profile in the vertical direction or through patterned convection states. Incompressibility adds a constraint that is easier to accommodate in direct solvers than in nonlinear Newton solvers. Figure~\ref{f:convection} demonstrates that the predictor-corrector algorithm applied with a direct solver resolves the bifurcation from conducting to convecting states well. We used the total area of the set $\{T>0.7(T_\mathrm{top}-T_\mathrm{bottom})\}$ as the feature function, which distinguishes well between conduction and convection, and the initial bifurcation point was found with a line search of this feature function in the interval $\mathrm{Gr}\in[2.4]$ for $\mathrm{Pr}=15$ as outlined in the initialization section in \S\ref{s:3}.}

\subsubsection*{Fold bifurcation curves}

Next, we consider fold bifurcations of spots and stripes in the planar Swift--Hohenberg equation posed on a square domain. For each of these two cases, we choose a single deterministic initial condition that consists of spots (or stripes) in the left half of the square domain and the homogeneous rest state in the right half of the domain. We apply the empirical measure $\mu_{f_\mathrm{RoundDistr},\mathrm{bag}}^{N}$ to the $\alpha$-shapes in the multiset of solution profiles to trace out the bifurcation curves.

\begin{figure}
\centering
\includegraphics{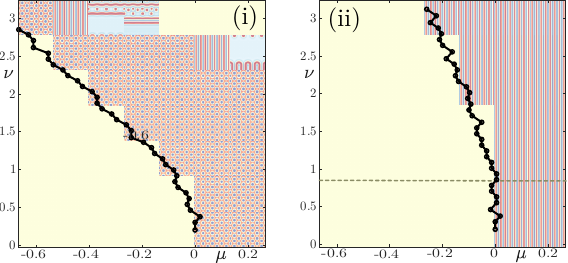}
\caption{We illustrate the computation of fold bifurcations of spots and stripes in the planar Swift--Hohenberg equation. Panel~(i) contains the fold curve of spots, which emanates from the origin. Panel~(ii) shows results for stripes \changed{(starting from one-dimensional initial conditions}): Below the line $\nu=\nu_*:=\sqrt{27/38}\approx0.843$, the curve reflects the supercritical bifurcation of stripes along $\mu=0$ into the region $\mu>0$ where the rest state $U=0$ is unstable. At $\nu=\nu_*$ (shown as the dotted horizontal line), this bifurcation becomes subcritical, and an additional bifurcation curve emerges at $(\mu,\nu)=(0,\nu_*)$ that corresponds to fold bifurcations of stripes.}
\label{f:folds}
\end{figure}

As shown in Figure~\ref{f:folds}(i), this approach accurately traces the fold bifurcation curve of spots that emerges from the origin in parameter space. Figure~\ref{f:folds}(ii) shows the results for stripes. For $0\leq\nu\leq\nu_*:=\sqrt{27/38}$, stripes bifurcate supercritically from the homogeneous rest state along the Turing bifurcation curve $\mu=0$. As $\nu$ crosses $\nu=\nu_*$, the bifurcation to stripe patterns along the Turing curve $\mu=0$ becomes subcritical, and a genuine fold bifurcation curve of stripe patterns emerges at $(\mu,\nu)=(0,\nu_*)$ and reaches into the region $\mu<0$. Since our prepared initial condition consists of stripes and the homogeneous rest state in the left and right halves of the square domain, our continuation framework first traces out the Turing curve $\mu=0$ before it picks up the fold bifurcation curve of stripes at $\nu=\nu_*$.

\subsubsection*{Comparison of feature functions}

\begin{figure}
\centering
\includegraphics[width=0.9\textwidth]{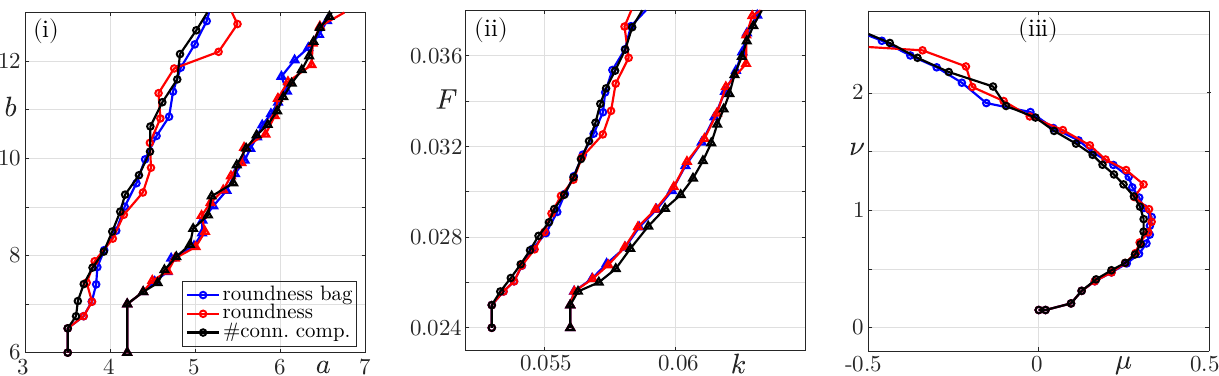}
\caption{
We compare the bifurcations curves for the (i) Brusselator, (ii) Gray--Scott, and (iii) Swift-Hohenberg models obtained through the pattern statistics $\mu_f^N$ for the number $f=f_\mathrm{Conn}\in\N$ of connected components in each pattern and the roundness score distribution $f=f_\mathrm{RoundDistr}\in\Prob([0,\mathfrak{m}])$ with the curves obtained using the bagged empirical roundness score $\mu_{f_\mathrm{RoundDistr},\mathrm{bag}}^{N}$.}
\label{f:comparison}
\end{figure}

So far, we used the bagged empirical pattern statistics $\mu_{f_\mathrm{RoundDistr},\mathrm{bag}}^{N}$ derived from the roundness score distribution. We illustrate now how the bifurcation curves obtained from different feature functions compare to each other. Figure~\ref{f:comparison} contains a comparison of the curves obtained from $\mu_{f_\mathrm{RoundDistr},\mathrm{bag}}^{N}$ in the preceding sections with the curves obtained using the empirical measures $\mu_f^N$ for the number $f=f_\mathrm{Conn}\in\N$ of connected components and the roundness score distribution $f=f_\mathrm{RoundDistr}\in\Prob([0,\mathfrak{m}])$. Overall, these curves agree well though the curve associated with the roundness score distribution shows more variability compared to the other two curves.

\subsubsection*{Instability curves of 1D source defects}

\begin{figure}
\centering
\includegraphics[scale=0.85]{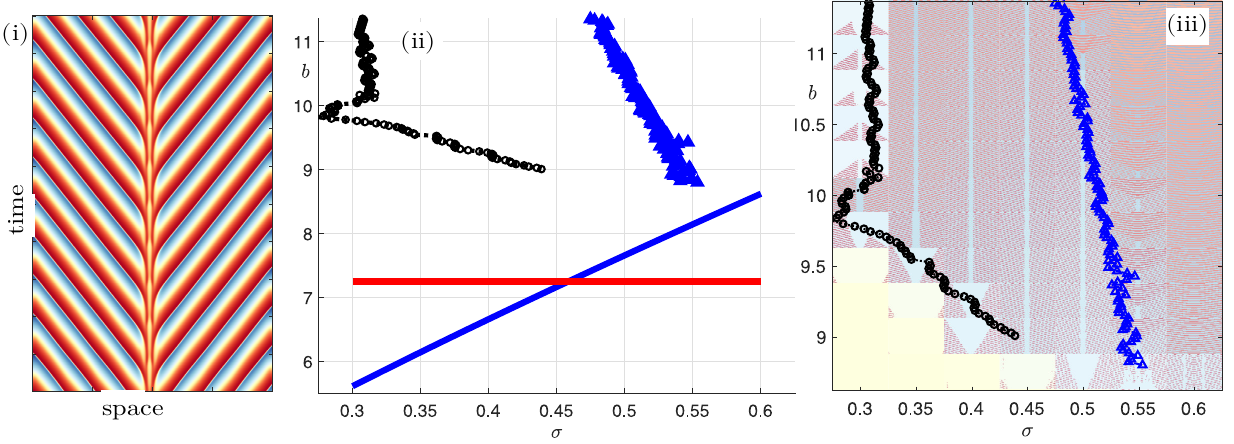}
\caption{\changed{Panel~(i) shows a space-time plot of a 1D source defect in the Brusselator model for the parameter values from \cite[Figure~3]{Perraud} with $\sigma:=D_1/D_2$. Panels~(ii) and~(iii) show instability curves of these defects together with the analytical Hopf (red) and Turing (blue) bifurcation curves of the homogeneous rest state in panel~(ii) and space-time plots shown in the inset tiles in panel~(iii).}}
\label{f:sourcedefect}
\end{figure}

\changed{In addition to the domain-filling pattern discussed earlier, the one-dimensional Brusselator model also exhibits complex spatio-temporal patterns that closely resemble experimental structures in the chlorite-iodide-malonic acid reaction \cite{Perraud}. One example are the source defects shown in Figure~\ref{f:sourcedefect}(i), which arise close to a codimension-two point where Hopf and Turing curves cross. Source defects are difficult to find analytically and to compute numerically, and we refer to \cite{SandstedeScheel2004} for theoretical results. The results in \cite{Tzou, Roberts} indicate that source defects may lie on complicated snaking branches similar to the simpler situation for localized roll patterns that we discuss further below. Here, we apply our algorithm with deterministic initial data to trace out two instability curves of source defects in the $(\sigma,b)$ parameter space; see Figure~\ref{f:sourcedefect}. We conjecture that these curves are the first fold curves in a complex snaking bifurcation diagram.}


\begin{figure}
\centering
\includegraphics[scale=1.25]{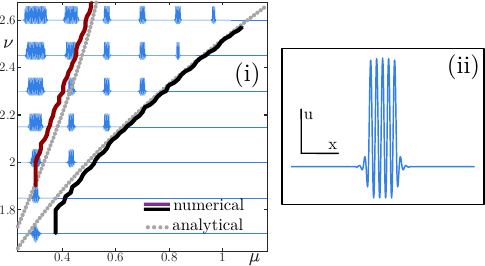}
\caption{Panel~(i) contains the numerically computed curves that delineate the snaking region in the one-dimensional Swift--Hohenberg equation together with fold bifurcation curves of localized roll patterns (a sample profile is shown in panel~(ii)) that were computed in AUTO. We also included direct simulations starting from localized roll patterns: in the upper left region, the roll plateau expands in time, while it shrinks to zero in the lower right region in parameter space.}
\label{f:snaking}
\end{figure}

\subsubsection*{Boundaries of snaking regions}

The one-dimensional Swift--Hohenberg equation exhibits stationary solutions whose spatial profiles consist of a spatially localized periodic plateau; see Figure~\ref{f:snaking}(ii) for a sample profile. These localized roll solutions exist in an open region in parameter space and, for each fixed parameter value $(\mu,\nu)$ inside this region, there are \changed{countably} many localized roll solutions that differ by the length $L\in\ell\N$ of the spatially periodic plateaus where $\ell>0$. The boundaries of the existence regions are delineated by bifurcation curves that correspond to folds of localized roll solutions (and we note that there are infinitely many of these fold bifurcation curves, namely one for each of the countable many localized rolls). We compute two of these fold bifurcation curves by selecting a single deterministic localized roll profile that has five maxima and use the number of connected components of the sublevel set $u^{-1}([\frac12\max u,\infty))$ measured by the feature function $f_\mathrm{Conn}$ of $\alpha$-shapes to trace out the bifurcation curve: note that this feature functions reflects the number of maxima of the solution after integrating in time. The result is shown in Figure~\ref{f:snaking}(i) and compared with AUTO computations of the same fold bifurcations. We conclude that our approach captures the snaking region well.


\subsubsection*{Agent-based models}

\begin{figure}
\centering
\includegraphics[width=0.75\textwidth]{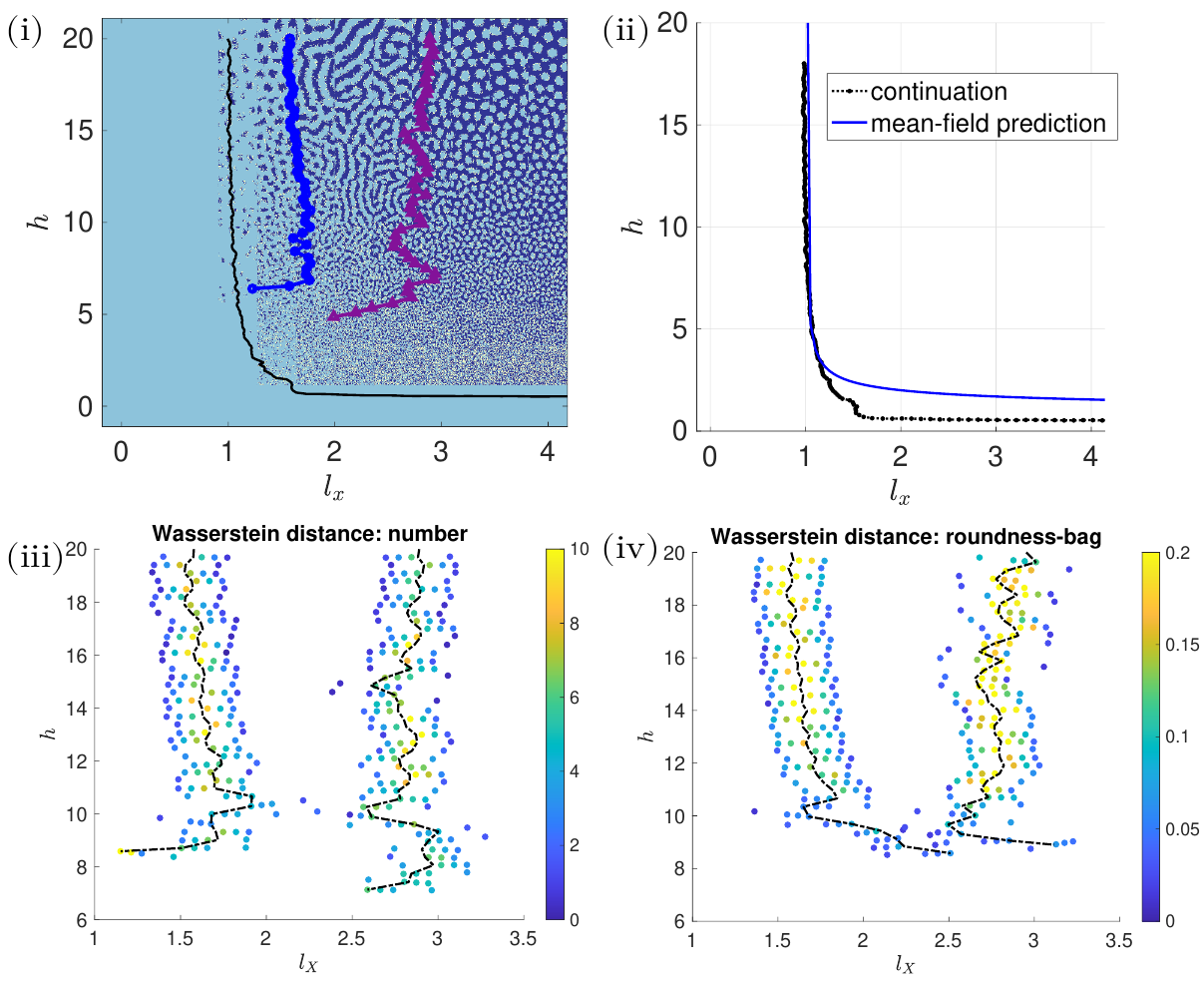}
\caption{\changed{Panel~(i) contains (1) the Turing curve in black, (2) the transition curve from melanophore spots to stripes in blue, and (3) the transitition curve from stripes to xanthophore spots in purple for the Bullara--De Decker model. The parameters $h$ and $l_x$ represent, respectively, the spatial extent and the strength of the long-range interaction mechanism that facilitates pattern formation. A comparison of the Turing curve with the mean-field prediction derived in \cite{Bullara} is shown in panel~(ii). As shown in panels~(iii) and~(iv), the Wasserstein distance of the pattern statistics across the two spot-stripe transition curves decreases very significantly as $h$ decreases below $10$: setting a stopping criterion based on a percentage of the initial Wasserstein distance at the start of the continuation would have terminated the continuation when the pattern statistics become indistinguishable.}}
\label{f:abm}
\end{figure}

\changed{To illustrate the applicability of our approach to stochastic many-particle systems, we report on an application to the Bullara--de~Decker model \cite{Bullara}, which is an agent-based lattice model designed to capture pigment patterns on zebrafish. The model focuses on three cell types (precursor cells, yellow xanthophores, and black melanophores), which interact with other over short and long spatial distances. It is inherently stochastic so traditional predictor-corrector algorithms are not applicable. In Figure~\ref{f:abm}, we show the results of our algorithm for (1) Turing bifurcations, (2) transitions from melanophore spots to stripes, and (3) transitions from stripes to xanthophore spots in the two parameters $(h,l_x)$ that enter into the model. The parameters $h$ and $l_x$ represent, respectively, the spatial extent and the strength of the mechanism that facilitates the transition from precursor cells to melanophores caused by long-range interactions with xanthophores. We note that mean-field theory \cite{Bullara} yields a formula for the Turing curve, but we are not aware of any previous computational or theoretical results for the spots-to-stripe transition curves. It is noticeable in Figure~\ref{f:abm}(i) that the transition curves begin to meander towards smaller values of $h$: this is expected since $h$ is linked to the spatial lattice wavelength of the patterned states in this model, so that small values of $h$ correspond to very fine-grained patterns. The continuation is terminated if it fails to detect an interval on which the average number of shapes differs by one: setting instead a condition based on the Wasserstein distance would have removed the bottom part of the two transition curves where the pattern statistics can no longer be distinguished; see Figure~\ref{f:abm}(iii)-(iv). The initial bifurcation points were found through a line search in $l_x$ for fixed $h=20$. To continue Turing bifurcations, we use $N=5$ simulations with a binary value that identifies whether the final pattern is homogeneous or not. For the transition curves, we use $N=15$ simulations with the number of $\alpha$-shapes ($\alpha=2$) of melanophores or xanthophores as the feature function. The key challenge for this model is the presence of outliers, and we needed to postprocess the output of the numerical simulations to remove these outliers: we follow the approach in \cite{zebrafish-tda, zebrafish-comparison} and remove the $10\%$ cells of each cell type whose average distance to the nearest $10$ cells of the same type is largest.}

\subsubsection*{Detection of bifurcations of spiral waves}

\begin{figure}
\centering
\includegraphics[scale=0.9]{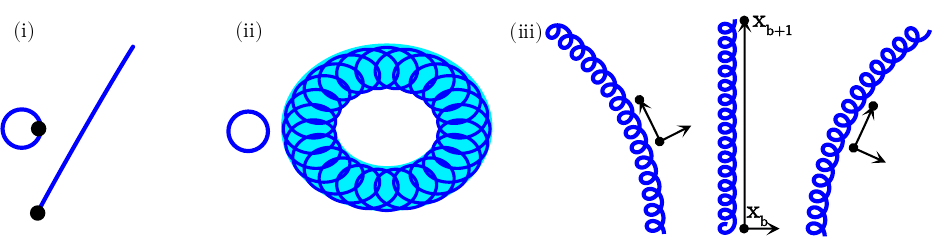}
\caption{Panel~(i) shows the tip trajectories of a rigidly-rotating spiral (circle) and a retracting spiral (line) with initial tip position indicated by a filled circle: the arc length of the tip trajectory can be used to distinguish these cases. Panel~(ii) illustrates the tip trajectories of rigidly-rotating and meandering spiral waves: here, the area of the face of the $\alpha$-shape can be used to distinguish these trajectories. Panel~(iii) shows the transition from inwardly meandering to drifting to outwardly meandering spiral waves: with the points $x_b$ and $x_{b+1}$ in the feature function $f_\mathrm{drift}$ defined in (\ref{e:ffd}) as indicated, the function $f_\mathrm{drift}$ will be zero for the drifting tip trajectory in the center, negative for the trajectory on the left and positive for the trajectory on the right.}
\label{f:tip_features}
\end{figure}

Next, we describe how our approach via feature functions can be used to trace out bifurcation curves of spiral waves. We refer to \cite[\S12 and Figure~12.8]{SandstedeScheel} and the references therein for background and more details on the bifurcations we consider here. Throughout, we use a single deterministic rigidly-rotating spiral-wave profile as the initial condition. We characterize spiral waves through the shape of their tip trajectories. The location $x(t)\in\R^2$ of the tip of a spiral wave $U(x,t)\in\R^2$ can be defined, for instance, through the requirement that $U(x(t),t)=\bar{U}$ for some fixed $\bar{U}\in\R^2$ (if $U\in\R^d$ with $d>2$, we define the tip position using two of the $d$ components of $U$). The tip location $x$ can be computed numerically using Newton's method applied to the equation $U(x,t)-\bar{U}=0$ at each time point $t$ during a direct numerical simulation. Having computed the time-dependent tip location $x(t)$, we define the tip trajectory $\mathcal{T}$ by $\mathcal{T}:=\{x(t)\colon t\in[0,T]\}$. The tip trajectory of a rigidly-rotating spiral wave is a circle, and we focus first on bifurcations where the shape of the spiral tip trajectory ceases to be a circle. The feature functions we choose to trace bifurcation curves will depend on the specific bifurcation scenario we are interested in, and we now discuss these choices in detail and refer to Figure~\ref{f:tip_features} for illustrations.

At \emph{retracting-wave bifurcations}, the temporal frequency of a rigidly-rotating spiral wave approaches zero, and the tip trajectory of the spiral wave changes from a circle to a semi-infinite line along which the spiral wave retracts to the domain boundary. We could therefore use the length $\tilde{f}_\mathrm{retract}(S_\alpha(\mathcal{T})):=\mathrm{Perimeter}(S_\alpha(\mathcal{T}))$ of the $\alpha$-shape $S_\alpha(\mathcal{T})$ of the tip trajectory as the feature function that distinguishes rigidly-rotating from retracting spiral waves. Alternatively, and this is the feature function we used in our numerical computations, we can use the Boolean function
\[
f_\mathrm{retract}(u) := 
\left\{ \begin{array}{lcl}
1 & & \max(u)-\min(u)<0.01 \\
0 & & \mathrm{otherwise,}
\end{array}\right.
\]
which detects transitions between the spiral wave ($f_\mathrm{retract}(u)=0$) and a homogeneous rest state ($f_\mathrm{retract}(u)=1$).

\emph{Meandering instabilities} correspond to Hopf bifurcations of rigidly-rotating spiral waves at which the tip trajectory acquires a second temporal frequency and changes from a circle to an epicycloid. In particular, the $\alpha$-shape $S_\alpha(\mathcal{T})$ of the spiral tip trajectory changes from a circle with vanishing area to an  annulus with strictly positive area. This motivates the feature function
\[
f_\mathrm{meander}(S_\alpha(\mathcal{T})) := \tanh\left(\frac{4\mu_\mathrm{Leb}(S_\alpha(\mathcal{T}))}{d_1d_2}\right)
\]
to detect transitions from rigidly-rotating to meandering spiral waves, where $d_j$ denotes the length of the interval $P_j S_\alpha(\mathcal{T})$ for the projection $P_j$ of $\R^2$ onto the $j$th component for $j=1,2$.

\emph{Drifting spiral waves} arise from meandering spiral waves when the tip trajectory becomes unbounded; see Figure~\ref{f:tip_features} for a sketch. Drifting spirals exist along curves in parameter space, and we continue these curves as follows. We compute the $\alpha$-shape $S_\alpha(\mathcal{T})$ of the spiral tip trajectory for a large value of $\alpha$, so that the $\alpha$-shape is close to the convex hull of $\mathcal{T}$. Denote by $\{x_j\}_{j=1,\ldots,s}$ the vertices of the polygon $S_\alpha(\mathcal{T})$, which we order so that $x_j$ and $x_{j+1}$ are adjacent on $S_\alpha(\mathcal{T})$, where we set $x_{s+1}:=x_1$. Let $1\leq b\leq s$ denote the index for which $|x_{b+1}-x_b|$ is largest. We then define the feature function
\begin{equation}\label{e:ffd}
f_\mathrm{drift}(S_\alpha(\mathcal{T})) := \frac{1}{s} \sum_{j=0}^{s} \sign\left\langle x_j-x_b, (x_{b+1}-x_b)^\perp\right\rangle
\end{equation}
to continue drifting spiral waves.

\begin{figure}
\centering
\includegraphics{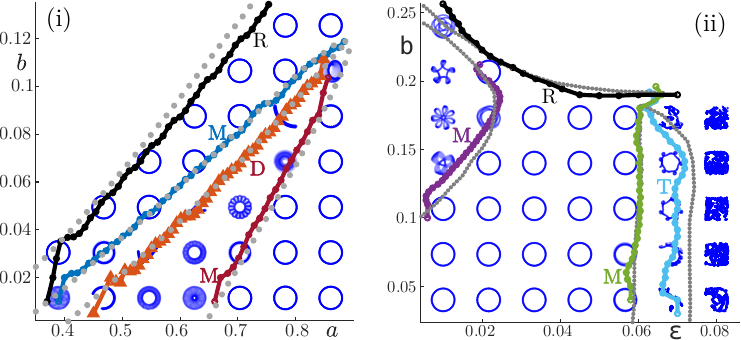}
\caption{Shown are retracting-wave bifurcation curves (R), curves corresponding to bifurcations from rigidly-rotating to meandering spirals (M), existence curves of drifting spiral waves (D), and transition curves from regular spiral waves to spiral-wave turbulence (T) for the Barkley model in panel~(i) and the B\"ar--Eiswirth model in panel~(ii). The tiles show the tip trajectories of spiral waves in the different regions of parameter space obtained from direct numerical simulations. Shown in gray dotted lines are the bifurcation curves obtained previously in \cite[Figure~1]{Barkley} for the Barkley model and in \cite[Figure~4]{BarEiswirth} for the B\"ar--Eiswirth model (reproduced via digitizing the original images).}
\label{f:BBE}
\end{figure}

\emph{Spiral-wave turbulence} arises when a large-scale spiral wave breaks up into many small spiral-wave segments. In this case, each spiral-wave segment has its own tip, and the equation $U_1(x,t)-\bar{U}=0$ that defines the tip will therefore have many solutions. In our algorithm, we apply Newton's method to the tip equation starting with initial data on a grid that spans the domain and collect the resulting tip positions in the set $X_t$ for each time $t$ during the direct simulation. Our feature function is then given by
\[
f_\mathrm{turbulence}(U_1) :=
\left\{ \begin{array}{lcl}
1 & & \mathrm{diam}(X_{t_\mathrm{max}})>10 \\
0 & & \mathrm{otherwise},
\end{array}\right. \quad
X_t := \{x\colon U(x,t)=\bar{U}\}, \quad
t_\mathrm{max} := \argmax_t \#\{x\in X_t\},
\]
which tests whether multiple tips occur as some time point during the simulation. Though the feature function $f_\mathrm{turbulence}$ is not continuous, it allows us to delineate transitions to turbulence efficiently and accurately.

We apply our continuation framework with the feature functions described above to the Barkley and the B\"ar--Eiswirth models, which exhibit retracting-wave instabilities, transitions to meandering and drifting spirals, and the emergence of spiral-wave turbulence. The results of our numerical continuation are shown in Figure~\ref{f:BBE}, where we also include the results of direct numerical simulations as well as the bifurcation curves obtained in \cite[Figure~1]{Barkley} and \cite[Figure~4]{BarEiswirth} for comparison.

\begin{figure}
\centering
\includegraphics[scale=1.2]{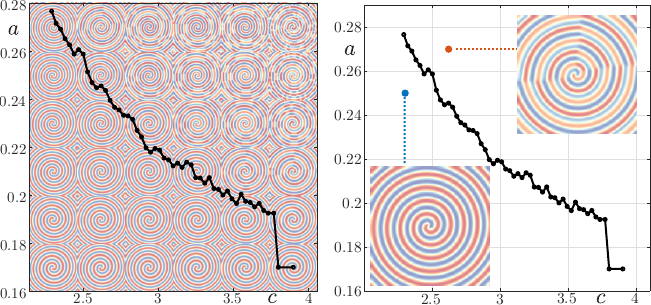}
\caption{Shown is the period-doubling curve of spiral waves for the R\"ossler system using the feature function $f_\mathrm{pd}$ with $k=3$. The tiles in the left panel correspond to the results of direct numerical simulations; the two insets in the right panel illustrate the difference between rigidly-rotating and period-doubled spiral waves.}
\label{f:rossler}
\end{figure}

\emph{Period-doubling bifurcations of spiral waves} lead to spiral waves with broken spiral-arm segments as shown in Figure~\ref{f:rossler}. These bifurcations cannot be detected easily by tip trajectories (while it is known that period-doubled spiral waves will drift, the drift speed is typically very small), and we therefore rely on a feature function that is based on the $\alpha$-shape of the full spiral profile. For a fixed value $1\leq k\leq d$, we select the function
\[
f_\mathrm{pd}(A) := \frac{1}{\beta(A)} \sum_{j=1}^{\beta(A)} \mu_\mathrm{Leb}(A_j), \qquad
A := U_k^{-1}([0.1\min U_k+0.9\max U_k,\infty))
\]
that reflects the sum of the areas of the $\beta(A)$ connected components of the set $U_k^{-1}([0.1\min U_k+0.9\max U_k,\infty))$, which consists of points $x$ for which $U_k(x)$ is close to $\max U_k$. Figure~\ref{f:rossler} shows the results of the continuation of period-doubling bifurcations of spiral waves in the three-component R\"ossler system for which we take $k=3$. The comparison with direct numerical simulations indicates that our feature functions traces out the bifurcation curve accurately.

\subsubsection*{Freezing method for spiral waves}

The freezing method was developed in \citep{Thuemmler1,Thuemmler2} to compute relative equilibria of reaction-diffusion systems via direct numerical simulations by adding algebraic constraints that keep the solution profile frozen at a specific location in space. The advantage of this approach is that it can be used to compute traveling waves by fixing their position and, at the same time, calculating their accumulated position (and velocity) without a prior knowledge of their speed; the disadvantage is that the method requires the use of solvers for algebraic-differential systems in order to account for the algebraic constraints that fix the location of the wave.

Here, we apply this approach to relative equilibria and relative periodic orbits in the form of rigidly-rotating as well as meandering and drifting spiral waves. The advantage of using the freezing method is that the spiral waves cannot leave the domain (which will happen for drifting spiral waves) and that this method directly generates the frequency and tip position of the underlying spiral wave. The approach we propose here relies on $\alpha$-shapes and does not require the use of solvers for algebraic-differential systems.

\begin{figure}
\centering
\includegraphics[scale=0.8]{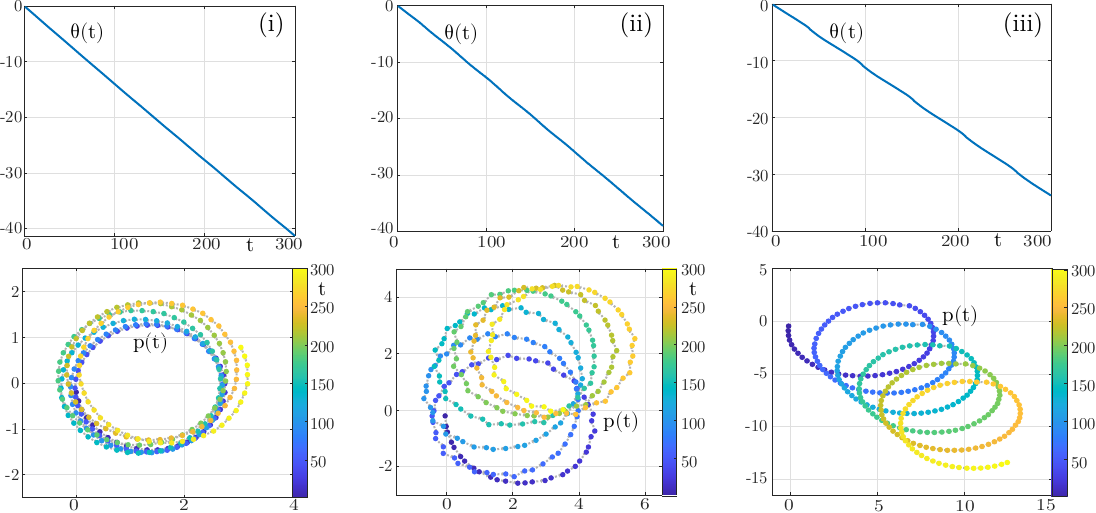}
\caption{Shown are the cumulative angles $\theta(t)$ (top row) and the tip positions $p(t)$ (bottom row) of rigidly-rotating, meandering, and drifting spiral waves in panels (i), (ii), and (iii), respectively. The spiral waves were computed using the implementation of the freezing method via $\alpha$-shapes for the Barkley model posed on a disk.}
\label{f:freezing}
\end{figure}

Given the initial condition $U(x,0)$ of the spiral-wave solution we plan to compute using the freezing method, we define $A(0):=S_\alpha(U_1^{-1}((-\infty,c]),0))$. We then integrate the underlying PDE from $t=0$ to $t=t_1$ and set $A(t_1):=S_\alpha(U_1^{-1}((-\infty,c]),t_1))$. Next, we solve
\[
(\varphi_1,q_1) := \argmax_{(\varphi,q)} \mathrm{Area}(\mathrm{Overlap}(R_\varphi(A(0)+q),A(t_1))), \qquad
R_\varphi := \begin{pmatrix} \cos\varphi & \sin\varphi \\ -\sin\varphi & \cos\varphi \end{pmatrix}
\]
by rotating and shifting the $\alpha$-shape at $t=t_1$ so that it aligns best with the $\alpha$-shape at $t=0$. After the optimal transformation is obtained, we apply the inverse of the alignment transformation to the spatial grid at time $t=0$ to obtain the grid $x^\mathrm{grid}_1:=R_{\varphi_1}^Tx^\mathrm{grid}-q_1$ on which the aligned solution at time $t=t_1$ is defined. We use linear interpolation to define the solution at time $t=t_1$ on the original grid $x^\mathrm{grid}$ and repeat the process by solving the PDE starting at time $t=t_1$ and solving until $t=t_2$, proceeding as above with $t_1$ replaced by $t_2$. At time $t=t_n$, we collected angles $(\varphi_j)_{j=1,\ldots,n}$ and translations $(q_j)_{j=1,\ldots,n}$, which we need to integrate in time to obtain the cumulative angles $\theta_n$ and tip positions $p_n$. This can be done by composing the inverses of the alignments sequentially, and we obtain
\[
\theta_n = \sum_{j=1}^n \varphi_j, \qquad
p_n = R(\varphi_n) q_n + R(\varphi_{n-1})R(\varphi_n)q_{n-1} + \ldots + R\left(\sum_{j=1}^n \varphi_j\right) q_1.
\]
We apply this approach to rigidly-rotating, meandering, and drifting spiral waves in the Barkley model posed on a disk of radius $30$. The results are shown in Figure~\ref{f:freezing}.


\section{Discussion} \label{s:5}

In this paper, we outlined a framework that allowed us to trace out bifurcation and transition curves of patterned states via feature functions and pattern statistics. The advantage of the approach discussed here is that we can use randomized initial conditions and compare the values of measure-valued feature functions using Wasserstein distances. The main disadvantages are (i) the lack of differentiability of the bifurcation functions and (ii) the current restriction to two-dimensional parameter spaces. We note that the lack of smoothness is inherent to our framework since we rely on Wasserstein distance in spaces of probability measures for which no differentiable structure exists. Extensions to higher-dimensional parameter spaces might be possible using multi-dimensional continuation algorithms as outlined in \cite{Allgower, Henderson} and the references in these papers.

\changed{We view unsupervised classification methods that are build, for instance, on the results of topological data analysis as complementary to our approach. While unsupervised learning aids in the classification of patterns and has been successfully used in the analysis of agent-based models \cite{zebrafish-tda}, Turing bifurcations (see the recent work \cite{harrington}), and clustering of different cell configurations in heterogeneous cell populations \cite{Bhaskar}, these approaches do not lend themselves to the accurate delineation of boundaries of the classification regions. For instance, the boundaries identified in \cite{Bhaskar} for heterogeneous cell populations consist of vertical and horizontal line segments that provide very rough approximations of the actual boundaries.}

There are several numerical parameters that affect the computational algorithm, including the feature function itself, the randomization, the spatial step sizes, the integration time, the threshold for the sublevel sets, and the radius used in the calculation of $\alpha$-shapes. \changed{While we provided some results in \S\ref{s:35} that assess the robustness of our algorithm under changes of these parameters,} we did not systematically explore the dependence on these parameters, and we also did not explore systematic ways to calibrate and optimize them.

Finally, we briefly discuss extensions of the framework we introduced in this paper. \changed{We see stochastic agent-based models (for instance, for the description of pigment patterns in zebrafish \cite{Bullara, zebrafish-tda} or of spatial aggregation patterns of heterogeneous cell populations such as epithelial cells during wound healing \cite{Bhaskar}) as an interesting area to which our framework could be applied. For PDEs, it would be interesting to use the proposed framework to numerically continue the contact defects that were found in the Brusselator \cite{Tzou} and analyzed in \cite{Roberts} without using core-farfield decompositions \cite{Avery, LloydScheel, Morrissey}, which are difficult to adapt to this case due to the stiffness of the Brusselator model and the logarithmic phase corrections in the solution profiles.} It should be possible to use feature functions and pattern statistics also to infer and identify parameters in simulations. If a single feature function is not able to distinguish patterns across the entire parameter space, it might be possible to use "majority votes" of several feature functions to correctly classify patterns and infer parameter values. Another potential application is to evaluate feature functions on solution trajectories to infer time dynamics from data similar to how we used tip trajectories to classify spiral-wave dynamics. Finally, pattern statistics could be useful to compare and fit models via parameter optimization; see \cite{zebrafish-comparison} for initial work that pursues this idea.


\appendix
\renewcommand{\thesection}{Appendix~\Alph{section}.}

\section{Models}

\paragraph{Barkley and B\"ar--Eiswirth models:}
Both models are examples of the Fitzhugh--Nagumo equation
\[
u_t = \Delta u + \frac{1}{\epsilon}u(1-u)\left(u-\frac{v+a}{b}\right), \qquad
v_t = g(u,v).
\]
Simulations for both models are carried out using Barkley's code \textsc{ezspiral} \cite{EZspiral}. The Barkley model \cite{Barkley, Barkley1995} is characterized by the nonlinearity $g(u,v)=u-v$. We use the spatial domain $[0,100]\times[0,100]$ discretized with $501$ points in each direction, choose $\epsilon=0.02$, and vary $(a,b)$. The B\"ar--Eiswirth model \cite{BarEiswirth} has the nonlinearity
\[
g(u,v) = h(u) - v,\qquad
h(u) =
\begin{cases}
0 &\quad u < \frac{1}{3}\\
1 - 6.75u(u-1)^2 &\quad u \in [\frac{1}{3},1]\\
1 & \quad u>1.
\end{cases}
\]
We use the domain $[0,50]\times[0,50]$ with $501$ mesh points in each direction, choose $a=0.84$, and vary $(b,\epsilon)$.

\paragraph{Brusselator:}
The Brusselator model \cite{Brusselator} is defined by
\[
u_t = D_1 \Delta u + a - (b+1)u + vu^2, \qquad
v_t = D_2 \Delta v + bu - vu^2.
\]
The homogeneous rest state $(u,v)=(a,b/a)$ undergoes a Turing bifurcation along the curve $b=(1+a\sqrt{D_1/D_2})^2$. For the two-dimensional case, we set $D_1=4$ and $D_2=32$, pose the equation on the domain $[0,50]\times[0,50]$ discretized with $50$ mesh points in each direction, and evolve the system in time with step size $dt=0.005$ until $T=100$. \changed{For the one-dimensional case, we use the parameter values in \cite[Figure~3]{Perraud} with $\sigma:=D_1/D_2$.}

\paragraph{\changed{Bullara--De~Decker:}}
\changed{We refer to \cite{Bullara} for a detailed description of this model and to \cite[Captions of Figures~1 and~2]{Bullara} for the parameter values we used.}

\paragraph{Gray--Scott:}
The Gray--Scott model \cite{GrayScott} is given by
\[
u_t = D_1 \Delta u + u v^2 + F(1-u), \qquad
v_t = D_2 \Delta v + u v^2 - (F+k) v.
\]
We pose this system on the domain $[0,2.5]\times [0,2.5]$ with diffusion constants $D_1=2\times 10^{-5}$ and $D_2=10^{-5}$. We use a uniform spatial grid $200\times 200$ mesh points and evolve the system with temporal step size $dt=1$ until $T=10,000$. For the initial data, we perturb from the homogeneous rest state $(u,v)=(1,0)$ by creating ten random spots of $0.0625$ in the domain.

\paragraph{\changed{Rayleigh--B\'enard:}}
\changed{The non-dimensional Rayleigh--B\'enard system with Boussinesq approximation \cite{GFD} is given by
\[
\frac{\partial u}{\partial t} +  u \frac{\partial u}{\partial x} + v \frac{\partial u}{\partial y} = - \nabla p +\frac{1}{\mathrm{Re}} \nabla^2  u, \qquad
\frac{\partial v}{\partial t} +  u \frac{\partial v}{\partial x} + v \frac{\partial v}{\partial y} = - \nabla p +\frac{1}{\mathrm{Re}} \nabla^2  v + \frac{\mathrm{Gr}}{\mathrm{Re}^2}  (T-T_0), \]
\[
\frac{\partial u}{\partial x} + \frac{\partial v}{\partial y} = 0,\qquad
\frac{\partial T}{\partial t} +u \frac{\partial T}{\partial x} + v \frac{\partial T}{\partial y} =  \frac{1}{\mathrm{Pe}} \nabla^2 T. 
\]
We use the system with Reynolds number fixed to $\mathrm{Re}=100$, spatial domain size $[0,0.82]\times [0,0.2]$ with a uniform grid of size $100\times 67$. The top wall temperature is set to 25, and bottom temperature is set to 70, and we use Dirichlet conditions for all other boundaries. Starting from the initial condition $(u,v)=0$ and $T=T_0=25$ everywhere, solutions are evolved with step size $0.01$ until $t=50$.
}

\paragraph{R\"ossler:}
The R\"ossler model \cite{Rossler, Kapral} is given by
\[
u_t = 0.4\, \Delta u - v - w, \qquad
v_t = 0.4\, \Delta v + u + av, \qquad
w_t = 0.4\, \Delta w + uw - cw + 0.2.
\]
We use the domain $[0,250]\times[0,250]$ discretized with $526$ mesh points in each direction.

\paragraph{Schnakenberg:}
The Schnakenberg model \cite{Schnakenberg} is given by
\[
u_t = D_1 \Delta u - u + u^2v + \frac{b-a}{2}, \qquad
v_t = D_2 \Delta v -u^2v + \frac{b+a}{2}.
\]
The homogeneous rest state $(u,v)=(b,\frac{a+b}{2b^2})$ undergoes a Turing bifurcation when $a=b^2\sqrt{\frac{D_1}{D_2}}\left(2+b\sqrt{\frac{D_1}{D_2}}\,\right)$. We set $D_1=0.005$ and $D_2=1$, pose the equation on the spatial domain $[0,4]\times[0,4]$, and discretize with $60$ mesh points in each spatial dimension. Initial conditions are given by the homogeneous rest state with a small amount of uniform noise added. Solutions are evolved with time step $dt=0.001$ until $T=500$.

\paragraph{Swift--Hohenberg:}
The Swift--Hohenberg equation \cite{SH} is given by
\[
u_t = -(1+\Delta)^2 u + \mu u + \nu u^2 - u^3.
\]
The homogeneous rest state $u=0$ undergoes a Turing bifurcation at $\mu=0$. We pose this equation on both $x\in\R$ (to investigate localized roll solutions) and $x\in\R^2$ (to study spots and stripes). When $x\in\R$, we chose the domain $[-32\pi,32\pi]$ with a uniform grid of 512 points. Solutions were computed using the time step $dt=0.1$ and evolved until $T=50$. To delineate the existence region of localized roll solutions, we chose the initial condition $u(x,0)=(\tanh(x+4\pi)-\tanh(x-4\pi))\cos(x)$. For $x\in\R^2$, we used the domain $[0,16\pi]\times[0,16\pi]$ with $128$ mesh points in each direction. Solutions are solved in time using an ETD scheme with time step $0.1$ and evolved until $T=500$. Randomized initial data are sampled from $U(0,1.5)$, while we use the prepared initial conditions
\[
u_\mathrm{spots}(x,0) = \frac{1}{6}\left(\cos(x_1)+\cos\left(\frac{x_1+\sqrt{3}x_2}{2}\right)+\cos\left(\frac{x_1-\sqrt{3}x_2}{2}\right)\right), \qquad
u_\mathrm{stripes}(x,0) = 0.2\cos(x_1)
\]
for spots and stripes, respectively.


\bibliographystyle{sandstede}
\bibliography{data_driven_bifurcation_continuation}

\end{document}